\documentclass[english,superscriptaddress,aps,pra,reprint]{revtex4-1}
\usepackage[T1]{fontenc}
\usepackage[latin9]{inputenc}
\usepackage{rotating}
\usepackage{booktabs}
\usepackage{mathrsfs}
\usepackage{bm}
\usepackage{multirow}
\usepackage{amsmath}
\usepackage{amssymb}
\usepackage{graphicx}

\makeatletter

\providecommand{\tabularnewline}{\\}

\usepackage{braket}
\makeatother

\usepackage{babel}
\begin{document}
\title{Determining the Superconducting Transition Temperatures of Liquids}
\author{Huiying Liu}
\affiliation{International Center for Quantum Materials, Peking University, Beijing
100871, China}
\author{Ying Yuan}
\affiliation{State Key Laboratory for Artificial Microstructure and Mesoscopic
Physics, and School of Physics, Peking University, Beijing 100871,
China}
\author{Donghao Liu}
\affiliation{International Center for Quantum Materials, Peking University, Beijing
100871, China}
\author{Xin-Zheng Li}
\affiliation{State Key Laboratory for Artificial Microstructure and Mesoscopic
Physics, and School of Physics, Peking University, Beijing 100871,
China}
\author{Junren Shi}
\email{junrenshi@pku.edu.cn}

\affiliation{International Center for Quantum Materials, Peking University, Beijing
100871, China}
\affiliation{Collaborative Innovation Center of Quantum Matter, Beijing 100871,
China}
\begin{abstract}
We develop a non-perturbative approach for calculating the superconducting
transition temperatures ($T_{\mathrm{c}}$'s) of liquids. The electron-electron
scattering amplitude induced by electron-phonon coupling (EPC), from
which an effective pairing interaction can be inferred, is related
to the fluctuation of the $T$-matrix of electron scattering induced
by ions. By applying the relation, EPC parameters can be extracted
from a path-integral molecular dynamics simulation. For determining
$T_{\mathrm{c}}$, the linearized Eliashberg equations are re-established
non-perturbatively. We apply the approach to estimate $T_{\mathrm{c}}$'s
of metallic hydrogen liquids. It indicates that metallic hydrogen
liquids in the pressure regime from $0.5$ to $1.5\mathrm{\,TPa}$
have $T_{\mathrm{c}}$'s well above their melting temperatures, therefore
are superconducting liquids.
\end{abstract}
\maketitle

\section{Introduction}

Mercury, the only metallic element which is a liquid under the ambient
conditions, happens to be the first superconductor ever discovered.
At a superconducting transition temperature ($T_{\mathrm{c}}$) of
$4.1\,\mathrm{K}$, however, it is frozen long before entering into
the superconducting state. As a matter of fact, all superconductors
discovered so far are solids. It seems improbable to find a superconducting
liquid. Recently, the possibility emerges with the report of a possible
observation of the Wigner-Huntington transition to metallic hydrogen~\citep{dias2017}.
Theoretically, it is predicted that hydrogen forms an atomic metal~\citep{mcmahon2012}
and has a relatively low melting temperature in the pressure regime
from $0.5$ to $1.5\,\mathrm{TPa}$~\citep{chen2013,geng2015}. On
the other hand, $T_{\mathrm{c}}$ predicted for the solid phase of
metallic hydrogen is much higher than the melting temperature~\citep{mcmahon2011}.
It raises an intriguing question: can a metallic hydrogen liquid be
superconducting?

A theoretical answer to the question would require developing a formalism
for predicting $T_{\mathrm{c}}$'s of liquids. For metallic hydrogen
liquids, Jaffe and Ashcroft present an estimate of $T_{\mathrm{c}}$
in the density range within $1.2\leq r_{\mathrm{s}}\leq1.6$~\citep{jaffe1981},
where $r_{\mathrm{s}}\equiv(3/4\pi n_{\mathrm{e}})^{1/3}/a_{\mathrm{B}}$
is the dimensionless density parameter with $n_{\mathrm{e}}$ being
the electron density and $a_{\mathrm{B}}$ the Bohr radius. The density
range is now believed not in the regime forming the atomic metal~\citep{mcmahon2012}.
Their formalism is based on a heuristic generalization of the conventional
electron-phonon coupling (EPC) theory~\citep{grimvall1981,giustino2017},
which is developed specifically for ordinary solids, relies on the
harmonic approximation of ionic motions, and is perturbative by nature.
For liquids, however, the harmonic approximation breaks down and there
is no apparent small parameter to facilitate a perturbative treatment.
The applicability of the conventional EPC theory is therefore questionable. 

It is desirable to build the EPC theory on a firmer ground, and seek
for a formalism with applicability extendable to liquids and other
unconventional systems such as anharmonic solids~\citep{borinaga2016,errea2013}.
With the advances of modern computation techniques, e.g., the \emph{ab
initio} path-integral molecular dynamics (PIMD) methods~\citep{marx1996,craig2004},
we are now at a much better position for applying such a formalism
and updating the calculation of metallic hydrogen liquids. More intriguingly,
the development would also give rise to a prospect of searching for
high-$T_{\mathrm{c}}$ EPC superconductors in unconventional systems.

In this paper, we develop a non-perturbative approach for calculating
$T_{\mathrm{c}}$'s of liquids. The central ingredient of our approach
is an exact relation between the electron-electron scattering amplitude
induced by EPC and the fluctuation of the $T$-matrix of electron
scattering induced by ions. The fluctuation can be evaluated with
a PIMD simulation, and an effective pairing interaction can be inferred
from the scattering amplitude. Our approach thus enables the evaluation
of EPC parameters from first principles for liquids. For determining
$T_{\mathrm{c}}$, we re-derive the Eliashberg equations in a non-perturbative
context. The approach is applied to investigate the superconductivity
of the liquid phase of metallic hydrogen. We find that metallic hydrogen
liquids in the pressure regime from $0.5$ to $1.5\mathrm{\,TPa}$
have $T_{\mathrm{c}}$'s well above their melting temperatures, therefore
are superconducting liquids. 

The remainder of the paper is organized as follows. In Sec.~\ref{sec:Theory},
we develop the theory of the superconductivity in liquids and general
systems. Main theoretical results are summarized in Sec.~\ref{subsec:Summary-of-Main},
and the proofs of these results are discussed in subsequent subsections.
Based on the theory, a numerical implementation for metallic hydrogen
is detailed in Sec.~\ref{sec:Numerical-Implementation-for}, with
main results summarized in Sec.~\ref{subsec:Results}. Finally, Sec.~\ref{sec:Summary}
is a summary.

\section{Theory\label{sec:Theory}}

\subsection{Summary of Main Results\label{subsec:Summary-of-Main}}

In this subsection, we summarize the main theoretical results of this
paper. They form the theoretical basis of calculating $T_{\mathrm{c}}$'s
of liquids. The proof of these results are presented in subsequent
subsections.

\subsubsection{Notations\label{subsec:Notations}}

In our formalism, we define two kinds of single-particle Green's functions
for electrons. $\mathcal{G}\left[\bm{R}(\tau)\right]$ is the Green's
function of an electron system subjected to the ionic field with respect
to a given ion configuration (trajectory) $\bm{R}(\tau)$:
\begin{multline}
\mathcal{G}\left[\bm{R}(\tau)\right]\left(\bm{r}\tau,\bm{r}^{\prime}\tau^{\prime}\right)=\\
-\mathrm{Tr}\left\{ \hat{T}_{\tau}\left[\hat{\rho}_{\mathrm{ei}}\left[\bm{R}(\tau)\right]\hat{\psi}_{\sigma}(\bm{r}\tau)\hat{\psi}_{\sigma}^{\dagger}(\bm{r}^{\prime}\tau^{\prime})\right]\right\} ,\label{eq:Ges}
\end{multline}
where $\bm{R}(\tau)\equiv\left\{ \bm{R}_{i}(\tau),i=1\dots N_{\mathrm{i}}\right\} $
is the short-hand notation of the trajectories of $N_{\mathrm{i}}$
ions with $\tau\in[0,\hbar\beta)$, $\beta\equiv1/k_{\mathrm{B}}T$
being the imaginary time arising in the Matsubara representation~\citep{fetter2003,mahan2000},
$\hat{\psi}_{\sigma}(\bm{r}\tau)$ and $\hat{\psi}_{\sigma}^{\dagger}(\bm{r}^{\prime}\tau^{\prime})$
are electron field operators, $\hat{\rho}_{\mathrm{ei}}\equiv Z_{\mathrm{ei}}^{-1}\hat{T}_{\tau}\exp[-(1/\hbar)\int_{0}^{\hbar\beta}d\tau(\hat{K}_{\mathrm{e}}+\hat{V}_{\mathrm{ei}}(\tau))]$
denotes the effective density matrix of the electron system with a
grand-canonical Hamiltonian $\hat{K}_{\mathrm{e}}$ and subjected
to a $\tau$-dependent ionic field $\hat{V}_{\mathrm{ei}}(\tau)$.
See Sec.~\ref{subsec:Exact-decomposition-of} for details. Due to
the presence of $\hat{V}_{\mathrm{ei}}(\tau)$ which breaks both the
spatial and temporal translational symmetries, the Green's function
is in general \emph{not }a function of ($\bm{r}-\bm{r}^{\prime}$,
$\tau-\tau^{\prime}$).

The \emph{physical} Green's function, which is denoted as $\bar{\mathcal{G}}$,
is obtained from $\mathcal{G}\left[\bm{R}(\tau)\right]$ after an
ensemble average over ion trajectories. See Sec.~\ref{subsec:Exact-decomposition-of}
for the definition of the ensemble average. For liquids, both the
spatial and the temporal translational symmetries are recovered after
the average. As a result, $\bar{\mathcal{G}}$ is a function of ($\bm{r}-\bm{r}^{\prime}$,
$\tau-\tau^{\prime}$). We define its Fourier transform as
\begin{multline}
\bar{\mathcal{G}}\left(\omega_{n},\bm{k}\right)=\\
\int\mathrm{d}\tau\int\mathrm{d}\bm{r}e^{\mathrm{i}\omega_{n}(\tau-\tau^{\prime})-\mathrm{i}\bm{k}\cdot(\bm{r}-\bm{r}^{\prime})}\bar{\mathcal{G}}\left(\bm{r}-\bm{r}^{\prime},\tau-\tau^{\prime}\right),
\end{multline}
where $\omega_{n}\equiv(2n+1)\pi/\hbar\beta$, $n\in Z$ is a Fermionic
Matsubara frequency and $\bm{k}$ is a wave-vector. Note that we distinguish
a function from its Fourier transform by their arguments {[}i.e.,
$(\bm{r}-\bm{r}^{\prime},\tau-\tau^{\prime})$ vs. $(\omega_{n},\bm{k})${]}. 

We adopt an abbreviated matrix notation for presenting our formalism.
A hatted symbol, e.g., $\hat{\mathcal{T}}$ in Eq.~(\ref{eq:TRtau}),
denotes a matrix, while $\mathcal{T}_{11^{\prime}}$ in Eq.~(\ref{eq:Gamma})
denotes an element of the matrix. The indices of matrix elements are
denoted by (decorated) numbers (e.g., $1$, $1^{\prime}$ or $\bar{1}$)
instead of usual alphabets. The indices refer to the set of parameters
labeling the basis of the matrix. We choose the basis in a particular
way such that the average (physical) Green's function $\bar{\mathcal{G}}$
is diagonal, i.e., $\left[\bar{\mathcal{G}}\right]_{11^{\prime}}=\bar{\mathcal{G}}_{1}\delta_{11^{\prime}}$.
For liquids, the index $1$ refers to a Matsubara frequency-wave vector
pair $(\omega_{n},\bm{k})$, and $1^{\prime}$ to $(\omega_{n^{\prime}},\bm{k}^{\prime})$,
and $\bar{\mathcal{G}}_{1}\equiv\bar{\mathcal{G}}\left(\omega_{n},\bm{k}\right)$,
$\delta_{11^{\prime}}\equiv\delta_{\omega_{n},\omega_{n^{\prime}}}\delta_{\bm{k},\bm{k}^{\prime}}$.

For liquids, which have both the temporal and the spatial translational
symmetries, the basis is just the plane-wave function $\varphi_{\omega_{n}\bm{k}}(\bm{r}\tau)=(\hbar\beta V)^{-1/2}\exp\left(-\mathrm{i}\omega_{n}\tau+\mathrm{i}\bm{k}\cdot\bm{r}\right)$,
where $V$ is the total volume of the system. In this case, matrix
indices refer to the pair of $(\omega_{n},\bm{k})$. With the notation,
a matrix element $\mathcal{T}(\bm{r}\tau,\bm{r}^{\prime}\tau^{\prime})\equiv\braket{\bm{r}\tau|\hat{\mathcal{T}}|\bm{r}^{\prime}\tau^{\prime}}$
can be expressed as:
\begin{align}
\mathcal{T}(\bm{r}\tau,\bm{r}^{\prime}\tau^{\prime}) & =\sum_{\omega_{n},\omega_{n^{\prime}},\bm{k},\bm{k}^{\prime}}\mathcal{T}_{\omega_{n}\bm{k},\omega_{n^{\prime}}\bm{k}^{\prime}}\varphi_{\omega_{n}\bm{k}}(\bm{r}\tau)\varphi_{\omega_{n^{\prime}}\bm{k}^{\prime}}^{\ast}(\bm{r}^{\prime}\tau^{\prime})\nonumber \\
 & \equiv\sum_{11^{\prime}}\mathcal{T}_{11^{\prime}}\varphi_{1}(\bm{r}\tau)\varphi_{1^{\prime}}^{\ast}(\bm{r}^{\prime}\tau^{\prime}),\label{eq:Trr}
\end{align}
where the summations over the indices are interpreted as
\begin{equation}
\sum_{1}\equiv\sum_{\omega_{n}}\sum_{\bm{k}}.
\end{equation}

For crystalline solids, the basis should be chosen as $\varphi_{\omega_{n}a\bm{k}}(\bm{r}\tau)=(\hbar\beta V)^{-1/2}\exp\left(-\mathrm{i}\omega_{n}\tau+\mathrm{i}\bm{k}\cdot\bm{r}\right)u_{a\bm{k}}(\bm{r})$,
where $u_{a\bm{k}}$ denotes the periodic part of a Bloch wave function
with a quasi-wave-vector $\bm{k}$ and a band index $a$. See Sec.~\ref{subsec:Reduction-of-the}
for the construction of Bloch wave functions. In this case, matrix
indices refer to $(\omega_{n},\bm{k},a)$. The abbreviated form of
Eq.~(\ref{eq:Trr}) is still valid with the new interpretation of
the indices.

For amorphous solids, one can nevertheless find a set of eigenfunctions
which diagonalize $\bar{\mathcal{G}}$. In this case, the indices
could in general be interpreted as the pair of a Matsubara frequency
and an index to the eigenfunctions.

An index with a bar (e.g., $\bar{1}$) refers to a basis which is
the time-reversal of the basis referred by the index without a bar.
For instance, for $1\rightarrow(\omega_{n},\bm{k})$, $\bar{1}$ refers
to $(-\omega_{n},-\bm{k})$.

\subsubsection{Effective Interaction Mediated by Ions\label{subsec:Effective-Interaction-Mediated}}

We first present a set of exact relations by which the effective interaction
mediated by ions can be determined. We adopt Matsubara's imaginary-time
formalism since we are dealing with a finite-temperature equilibrium
problem~\citep{fetter2003,mahan2000}. 

The first equation determines the ion-induced scattering amplitude
of a pair of electrons (a Cooper pair) with state indices $1\equiv(\omega_{1},\bm{k}_{1})$
and $\bar{1}\equiv(-\omega_{1},-\bm{k}_{1})$ scattered to $1^{\prime}$
and $\bar{1^{\prime}}$, respectively:
\begin{equation}
\Gamma_{11^{\prime}}=-\beta\left\langle \left|\mathcal{T}_{11^{\prime}}\left[\bm{R}(\tau)\right]\right|^{2}\right\rangle _{\mathrm{C}},\label{eq:Gamma}
\end{equation}
where $\Gamma_{11^{\prime}}$ denotes the pair scattering amplitude,
and $\mathcal{T}_{11^{\prime}}\left[\bm{R}(\tau)\right]$ is the $T$-matrix
element of electron scattering from $1$ to $1^{\prime}$ induced
by the $\tau$-dependent ionic field with respect to $\bm{R}(\tau)$.
The average $\left\langle \dots\right\rangle _{\mathrm{C}}$ is over
the trajectories of ions in a classical ensemble isomorphic to the
original quantum ionic system (see Sec.~\ref{subsec:Exact-decomposition-of}),
and can be evaluated in, e.g., a PIMD simulation. 

The second one is the Lippmann-Schwinger equation which determines
the $T$-matrix:
\begin{equation}
\hat{\mathcal{T}}\left[\bm{R}(\tau)\right]=\hat{\mathcal{V}}\left[\bm{R}(\tau)\right]+\frac{1}{\hbar}\hat{\mathcal{V}}\left[\bm{R}(\tau)\right]\hat{\bar{\mathcal{G}}}\hat{\mathcal{T}}\left[\bm{R}(\tau)\right],\label{eq:TRtau}
\end{equation}
where $\bar{\mathcal{G}}$ denotes the temperature Green's function~\citep{fetter2003}
of electrons in the normal state of the liquid, and $\mathcal{V}\left[\bm{R}(\tau)\right]\equiv V_{\mathrm{ei}}\left[\bm{R}(\tau)\right]-\bar{\Sigma}$
is the scattering potential with $V_{\mathrm{ei}}\left[\bm{R}(\tau)\right]$
being the time-dependent ionic field with respect to $\bm{R}(\tau)$
and $\bar{\Sigma}$ being the self-energy with respect to $\bar{\mathcal{G}}$.
We note that the scattering is relative to an effective medium defined
by $\bar{\mathcal{G}}$, and as a result, $\langle\hat{\mathcal{T}}[\bm{R}(\tau)]\rangle_{\mathrm{C}}=0$.
We further note that $\bar{\mathcal{G}}=\langle\mathcal{G}[\bm{R}(\tau)]\rangle_{\mathrm{C}}$,
where $\mathcal{G}[\bm{R}(\tau)]$ is the temperature Green's function
of electrons subjected to $V_{\mathrm{ei}}\left[\bm{R}(\tau)\right]$.

Finally, the effective pairing interaction $\hat{W}$, which enters
into the linearized Eliashberg equations (see Sec.~\ref{subsec:Linearized-Eliashberg-Equations})
and determines $T_{\mathrm{c}}$, can be inferred from the pair scattering
amplitudes by solving a Bethe-Salpeter (BS) equation:
\begin{equation}
W_{11^{\prime}}=\Gamma_{11^{\prime}}+\frac{1}{\hbar^{2}\beta}\sum_{2}W_{12}\left|\bar{\mathcal{G}}_{2}\right|^{2}\Gamma_{21^{\prime}}.\label{eq:W}
\end{equation}

The three equations (\ref{eq:Gamma}--\ref{eq:W}) form the theoretical
basis of determining EPC for liquids. The applicability of the formalism
can be extended to general systems by properly interpreting the state
indices as indicated in Sec.~\ref{subsec:Notations}. We can show
that the conventional EPC formalism~\citep{grimvall1981,giustino2017}
is just a limiting form of our formalism. See Sec.~\ref{subsec:Reduction-of-the}.

\subsubsection{Linearized Eliashberg Equations\label{subsec:Linearized-Eliashberg-Equations}}

After obtaining the effective pairing interaction $\hat{W}$, we still
need a formalism for determining $T_{\mathrm{c}}$. In the conventional
Eliashberg theory, $T_{\mathrm{c}}$ is determined by solving the
linearized Eliashberg equations~\citep{rainer1974,allen1975,carbotte1990}:
\begin{align}
\rho\Delta_{n} & =\sum_{n^{\prime}}\left[\lambda(n^{\prime}-n)-\mu^{\ast}-\frac{\hbar\beta}{\pi}\left|\tilde{\omega}(n)\right|\delta_{nn^{\prime}}\right]\Delta_{n^{\prime}},\label{eq:Eliashberg1}\\
\tilde{\omega}(n) & =\frac{\pi}{\hbar\beta}\left(2n+1+\lambda(0)+2\sum_{m=1}^{n}\lambda(m)\right),n\ge0\label{eq:on}
\end{align}
and $|\tilde{\omega}(-n)|=|\tilde{\omega}(n-1)|$. A positive eigenvalue
$\rho$ indicates an instability toward forming Cooper pairs and the
superconducting state. The interaction parameters are determined by:
\begin{equation}
\lambda(n^{\prime}-n)=-\sum_{\bm{k}^{\prime}}W_{\bm{k}^{\prime}\bm{k}}(\omega_{n^{\prime}}-\omega_{n})\delta(\tilde{\epsilon}_{\bm{k}^{\prime}}-\mu),\label{eq:lambda}
\end{equation}
where $W_{\bm{k}^{\prime}\bm{k}}(\omega_{n^{\prime}}-\omega_{n})\equiv W_{1^{\prime}1}$
with $1\equiv(\omega_{n},\bm{k})$ and $1^{\prime}\equiv(\omega_{n^{\prime}},\bm{k}^{\prime})$
is assumed to be a function of $\omega_{n^{\prime}}-\omega_{n}$,
and $\tilde{\epsilon}_{\bm{k}^{\prime}}$ is the electron dispersion
renormalized by the real part of $\bar{\Sigma}$. In the conventional
theory, the Eliashberg equations are established in a perturbative
context by assuming that the vibration amplitudes of ions are small.
The assumption is obviously not valid for liquids.

Our conclusion, simply put, is that one can still apply the Eliashberg
equations to determine $T_{\mathrm{c}}$'s for liquids and general
systems. We can re-establish the Eliashberg equations without resorting
to the perturbative approach. In our context, however, we have to
interpret them differently. Equation~(\ref{eq:Eliashberg1}) is now
interpreted as the equation determining the instability toward forming
the superconducting states. On the other hand, Eq.~(\ref{eq:on})
is the result of the self-energy equation
\begin{equation}
\mathrm{Im}\bar{\Sigma}_{1}=-\frac{1}{\hbar\beta}\sum_{1^{\prime}}W_{1^{\prime}1}\mathrm{Im}\bar{\mathcal{G}}_{1^{\prime}}\label{eq:imsigma}
\end{equation}
which is now interpreted as a generalized optical theorem~\citep{vanoosten1985}.
The proofs of these points are shown in Sec.~\ref{subsec:Linearized-Eliashberg-equation}. 

\subsection{Proofs}

To prove the main results outlined in the last subsection, we first
introduce two useful theoretical apparatuses, namely, the effective
action theory (Sec.~\ref{subsec:Effective-action-theory}) and the
exact decomposition of an electron-ion coupled system (Sec.~\ref{subsec:Exact-decomposition-of}).
Based upon these preparations, the main results are established in
Sec.~\ref{subsec:Effective-Pairing-Interaction} and \ref{subsec:Linearized-Eliashberg-equation}.
In Sec.~\ref{subsec:Reduction-of-the}, we further show that our
formalism is reduced to the conventional one when applied to ordinary
solids.

\subsubsection{Effective Action Theory\label{subsec:Effective-action-theory}}

The density functional theory (DFT) dictates that the ground state
energy (or grand potential) of an interacting quantum system is a
functional of density. The insight gives rise to a general framework
for treating interacting systems non-perturbatively. The theory could
be formally generalized to define a grand potential as a functional
of the Green's function. This is useful when single-particle excitations
are of interest. The construction is shown as follows.

The partition function of a general system, under the functional-integral
formalism, can be determined by~\citep{negele1988}
\begin{equation}
Z=\int\limits _{\psi(\hbar\beta)=-\psi(0)}\mathrm{D}\left[\psi^{\ast},\psi\right]\exp\left(-\frac{S\left[\psi,\psi^{\ast}\right]}{\hbar}\right),
\end{equation}

\begin{multline}
S\left[\psi,\psi^{\ast}\right]\equiv\int_{0}^{\hbar\beta}\mathrm{d}\tau\left[\psi^{\ast}(\tau)\cdot\left(\hbar\partial_{\tau}-\mu\right)\psi(\tau)\right.\\
\left.+K\left(\psi^{\ast}(\tau),\psi(\tau)\right)\right],
\end{multline}
where we assume that particles are Fermions, and $\psi$ denotes a
Grassmann field which fulfills the anti-periodic boundary condition
along the direction of the imaginary time: $\psi(\hbar\beta)=-\psi(0)$.
For brevity, we do not show explicitly the spatial dependence of the
field.

\paragraph{Normal systems}

We then introduce an auxiliary field $J(\bm{r}^{\prime}\tau^{\prime},\bm{r}\tau)=\sum_{1}J_{1}\varphi_{1}(\bm{r}^{\prime}\tau^{\prime})\varphi_{1}^{\ast}(\bm{r}\tau)$
which conjugates to the Green's function and modifies the action by:
\begin{align}
S_{J}\left[\psi,\psi^{\ast}\right]= & S\left[\psi,\psi^{\ast}\right]-\int\mathrm{d}\tau\mathrm{d}\tau^{\prime}\int\mathrm{d}\bm{r}\mathrm{d}\bm{r}^{\prime}\nonumber \\
 & \times J(\bm{r}^{\prime}\tau^{\prime},\bm{r}\tau)\psi(\bm{r}\tau)\psi^{\ast}(\bm{r}^{\prime}\tau^{\prime})\\
= & S-\sum_{1}J_{1}\psi_{1}\psi_{1}^{\ast},
\end{align}
where $\psi_{1}\equiv\int\mathrm{d}\tau\int\mathrm{d}\bm{r}\varphi_{1}^{\ast}(\bm{r}\tau)\psi(\bm{r}\tau)$
with $\varphi_{1}(\bm{r}\tau)$ being the basis function defined in
Sec.~\ref{subsec:Notations}.

With $S_{J}$, we can define a partition functional $Z[J]$. The temperature
Green's function in the presence of $J$ can be determined by a functional
derivative:
\begin{equation}
\mathcal{G}_{1}[J]=-\hbar\frac{\delta\ln Z[J]}{\delta J_{1}},
\end{equation}
according to the definition of the Green's function. The relation
basically maps $J$ to $\mathcal{G}$.

By assuming the map from $J$ to $\mathcal{G}$ is invertible, we
can define a grand potential as a functional of $\mathcal{G}$ by
applying the Legendre transformation:
\begin{align}
\Omega\left[\mathcal{G}\right] & =-\frac{1}{\beta}\ln Z\left[J\right]-\frac{1}{\hbar\beta}\sum_{1}J_{1}\mathcal{G}_{1}\nonumber \\
 & \equiv-\frac{1}{\beta}\ln Z\left[J\right]-\frac{1}{\hbar\beta}\mathrm{Tr}\hat{J}\hat{\mathcal{G}}.
\end{align}
With the grand potential functional, the Green's function can be obtained
by solving the equation
\begin{equation}
\hbar\beta\frac{\delta\Omega\left[\mathcal{G}\right]}{\delta\mathcal{G}_{1}}=-J_{1}.\label{eq:variational}
\end{equation}
It becomes a variational principle when $J=0$.

Following the procedure, it is not difficult to construct the functional
for a non-interacting system~\citep{kotliar2006}: 
\begin{equation}
\beta\Omega_{0}[\mathcal{G}]=\mathrm{Tr}\ln\hat{\mathcal{G}}-\mathrm{Tr}[\hat{\mathcal{G}}_{0}^{-1}\hat{\mathcal{G}}-I],
\end{equation}
with $\hat{\mathcal{G}}_{0}^{-1}\equiv[-\partial_{\tau}+\mu/\hbar+(\hbar/2m)\nabla^{2}]\delta(\tau-\tau^{\prime})\delta(\bm{r}-\bm{r}^{\prime})$.

For an interacting system, one can decompose the grand potential functional
into two parts:
\begin{equation}
\Omega\left[\mathcal{G}\right]=\Omega_{0}\left[\mathcal{G}\right]+\Omega_{\mathrm{LW}}\left[\mathcal{G}\right],
\end{equation}
where $\Omega_{\mathrm{LW}}\left[\mathcal{G}\right]$ is called Luttinger-Ward
functional which accounts for interaction effects~\citep{luttinger1960}.
With the Luttinger-Ward functional, we can define a self-energy functional
\begin{equation}
\Sigma\left[\mathcal{G}\right]=-\hbar\beta\frac{\delta\Omega_{\mathrm{LW}}\left[\mathcal{G}\right]}{\delta\mathcal{G}}.
\end{equation}
By applying Eq.~(\ref{eq:variational}), we obtain a self-consistent
Dyson equation for determining $\mathcal{G}$:
\begin{equation}
\left\{ \hat{\mathcal{G}}_{0}^{-1}-\frac{\hat{J}+\hat{\Sigma}\left[\mathcal{G}\right]}{\hbar}\right\} \hat{\mathcal{G}}=I.\label{eq:Dyson}
\end{equation}
We note that the equation is formally exact provided that the functional
form of the self-energy is known.

More generally, we can introduce an auxiliary field $J(\bm{r}^{\prime}\tau^{\prime},\bm{r}\tau)=\sum_{1}J_{1^{\prime}1}\varphi_{1^{\prime}}(\bm{r}^{\prime}\tau^{\prime})\varphi_{1}^{\ast}(\bm{r}\tau)$
which is non-diagonal in the basis. In this case, we can also define
a grand potential functional $\Omega\left[\mathcal{G}\right]$ without
assuming $\hat{\mathcal{G}}$ to be diagonal. For this case, the counterpart
of Eq.~(\ref{eq:variational}) is
\begin{equation}
\hbar\beta\frac{\delta\Omega\left[\mathcal{G}\right]}{\delta\mathcal{G}_{11^{\prime}}}=-J_{1^{\prime}1}.
\end{equation}

\paragraph{Superconducting systems}

For treating superconducting systems, it is necessary to further generalize
the formalism. This is to replace the Green's function $\mathcal{G}$
with a $2\times2$ matrix of Green's functions in the Nambu representation~\citep{scalapino1969}:
\begin{equation}
\mathscr{G}_{1}=\left[\begin{array}{cc}
\mathcal{G}_{1} & \mathcal{F}_{1}\\
\mathcal{F}_{1}^{\ast} & -\mathcal{G}_{\bar{1}}
\end{array}\right],
\end{equation}
where we introduce an anomalous Green's function $\mathcal{F}(\bm{r}\tau,\bm{r}^{\prime}\tau^{\prime})=-\langle\hat{T}_{\tau}\hat{\psi}_{\uparrow}(\bm{r}\tau)\hat{\psi}_{\downarrow}(\bm{r}^{\prime}\tau^{\prime})\rangle$~\citep{fetter2003}
with the subscripts of the field operators indexing spin components.
By introducing an auxiliary field $\Delta(\bm{r}^{\prime}\tau^{\prime},\bm{r}\tau)=\sum_{1}\Delta_{1}\varphi_{\bar{1}}(\bm{r}^{\prime}\tau^{\prime})\varphi_{1}(\bm{r}\tau)$
conjugated to $\mathcal{F}$, we have:
\begin{align}
S_{\Delta}= & S-\int\mathrm{d}\tau\mathrm{d}\tau^{\prime}\int\mathrm{d}\bm{r}\mathrm{d}\bm{r}^{\prime}\nonumber \\
 & \times\left[\Delta^{\ast}(\bm{r}^{\prime}\tau^{\prime},\bm{r}\tau)\psi_{\uparrow}(\bm{r}\tau)\psi_{\downarrow}(\bm{r}^{\prime}\tau^{\prime})+\mathrm{h.c.}\right]\\
= & S-\sum_{1}\left(\Delta_{1}^{\ast}\psi_{1\uparrow}\psi_{\bar{1}\downarrow}+\mathrm{h.c.}\right).\label{eq:DSD}
\end{align}
It is not difficult to repeat the above discussions to define a grand
potential functional $\Omega\left[\mathcal{G},\mathcal{F}\right]$.
In addition to Eq.~(\ref{eq:variational}), we have:
\begin{equation}
\hbar\beta\frac{\delta\Omega\left[\mathcal{G},\mathcal{F}\right]}{\delta\mathcal{F}_{1}}=-\Delta_{1}^{\ast}.\label{eq:DODF}
\end{equation}
The functional of the non-interacting reference system becomes:
\begin{equation}
\beta\Omega_{0}[\mathcal{G},\mathcal{F}]=\mathrm{Tr}\ln\hat{\mathscr{G}}-\mathrm{Tr}[\hat{\mathscr{G}}_{0}^{-1}\hat{\mathscr{G}}-I],
\end{equation}
where $\hat{\mathscr{G}}_{0}^{-1}\equiv[-\partial_{\tau}+\hbar^{-1}(\mu+(\hbar^{2}/2m)\nabla^{2})\hat{\tau}_{3}]\delta(\tau-\tau^{\prime})\delta(\bm{r}-\bm{r}^{\prime})$
with $\hat{\tau}_{3}$ being the third component of the Pauli matrices.

\paragraph{Functional expansion, stiffness theorem, and anomalous response function}

We exploit the fact that when the temperature approaches $T_{\mathrm{c}}$,
the amplitude of $\mathcal{F}$ must be small. As a result, we can
expand the functional as a Taylor series of $\mathcal{F}$. To the
second order, the expansion has the form:
\begin{multline}
\Omega\left[\mathcal{G},\mathcal{F}\right]=\Omega_{0}\left[\mathcal{G},\mathcal{F}\right]+\Omega_{\mathrm{LW}}^{\mathrm{N}}\left[\mathcal{G}\right]\\
+\frac{1}{(\hbar\beta)^{2}}\sum_{11^{\prime}}\mathcal{F}_{1}^{\ast}W_{11^{\prime}}\mathcal{F}_{1^{\prime}}+\dots\label{eq:Omegaexpansion}
\end{multline}
where $\Omega_{\mathrm{LW}}^{\mathrm{N}}\left[\mathcal{G}\right]\equiv\Omega_{\mathrm{LW}}\left[\mathcal{G},\mathcal{F}\rightarrow0\right]$
is the Luttinger-Ward functional for the normal state. The coefficients
are interpreted as the effective pairing interaction, and determined
by:
\begin{equation}
W_{11^{\prime}}=(\hbar\beta)^{2}\left.\frac{\delta^{2}(\Omega-\Omega_{0})}{\delta\mathcal{F}_{1}^{\ast}\delta\mathcal{F}_{1^{\prime}}}\right|_{\mathcal{F}\rightarrow0}.
\end{equation}

By applying Eq.~(\ref{eq:DODF}), we have:
\begin{equation}
\hbar\beta\left.\frac{\delta^{2}\Omega}{\delta\mathcal{F}_{1}^{\ast}\delta\mathcal{F}_{1^{\prime}}}\right|_{\mathcal{F}\rightarrow0}=-\left.\frac{\delta\Delta_{1}}{\delta\mathcal{F}_{1^{\prime}}}\right|_{\mathcal{F}\rightarrow0}\equiv-\left[\hat{\chi}^{-1}\right]_{11^{\prime}},\label{eq:stiffness}
\end{equation}
where we define an anomalous density response function
\begin{equation}
\chi_{11^{\prime}}=\left.\frac{\delta\mathcal{F}_{1}}{\delta\Delta_{1^{\prime}}}\right|_{\Delta\rightarrow0}
\end{equation}
which is just the matrix inverse of $\left[\delta\Delta_{1}/\delta\mathcal{F}_{1^{\prime}}\right]$.
Equation (\ref{eq:stiffness}) is nothing but the stiffness theorem
which could be established in the more general context~\citep{giuliani2005}. 

Combining these relations, we have
\begin{equation}
\hat{W}=\hbar\beta\left(\hat{\chi}_{0}^{-1}-\hat{\chi}^{-1}\right),\label{eq:Wchi}
\end{equation}
where $[\hat{\chi}_{0}]_{11^{\prime}}=-\hbar^{-1}\left|\bar{\mathcal{G}}_{1}\right|^{2}\delta_{11^{\prime}}$
is the anomalous response function for the non-interacting reference
system with respect to $\Omega_{0}\left[\mathcal{G},\mathcal{F}\right]$.

The anomalous response function can be related to a correlation function
in the functional integral formalism. We have:
\begin{align}
\mathcal{F}_{1} & =-\frac{1}{Z_{\Delta}}\int\mathrm{D}\left[\psi,\psi^{\ast}\right]\psi_{1\uparrow}\psi_{\bar{1}\downarrow}e^{-S_{\Delta}/\hbar},\\
\chi_{11^{\prime}} & =-\frac{1}{\hbar}\left\langle \left(\psi_{1\uparrow}\psi_{\bar{1}\downarrow}-\mathcal{F}_{1}\right)\left(\psi_{1^{\prime}\uparrow}\psi_{\bar{1^{\prime}}\downarrow}-\mathcal{F}_{1^{\prime}}\right)^{\ast}\right\rangle ,\label{eq:chi11}
\end{align}
where the average $\langle\dots\rangle\equiv Z^{-1}\int\mathrm{D}[\psi,\psi^{\ast}]\dots\exp(-S/\hbar)$.

\paragraph{Kohn-Sham decomposition}

With the formalism, we have a formal framework for treating many-body
physics non-perturbatively. The formalism is useful only when we know
the form of the functional. In real calculations, it is necessary
to adopt an approximation for the functional form. A sensible starting
approximation is based on the Kohn-Sham decomposition, by which the
Green's function is expressed in terms of Kohn-Sham wave-functions
and eigen-energies just like a non-interacting system. The approach
is then reduced to the ordinary Kohn-Sham theory. See Ref.~\citep{kotliar2006}
for more information. For the EPC of a system which is not regarded
as ``strongly correlated'', the approximation is usually adequate.
Actually, most modern-day first-principles calculations of EPC for
ordinary solids are based on the same approximation~\citep{giustino2017}.

\subsubsection{Exact Decomposition of An Electron-Ion Coupled System\label{subsec:Exact-decomposition-of}}

To treat a system involving strongly coupled electrons and ions, we
adopt an exact decomposition which separates the treatments of the
ion and electron degrees of freedom. The ion degrees of freedom can
be simulated by the PIMD. The electron subsystem is then mapped into
a system subjected to a stochastic time-dependent ionic field sampled
by the PIMD.

\begin{widetext}

The grand-canonical Hamiltonian of an electron-ion coupled system
can be in general written as (i.e., ``the Hamiltonian of everything''):

\begin{multline}
\hat{K}=\underset{\hat{K}_{\mathrm{e}}}{\underbrace{\sum_{\sigma}\int\mathrm{d}\bm{r}\hat{\psi}_{\sigma}^{\dagger}(\bm{r})\left[-\frac{\hbar^{2}}{2m_{\mathrm{e}}}\nabla_{\bm{r}}^{2}-\mu\right]\hat{\psi}_{\sigma}(\bm{r})+\frac{1}{2}\sum_{\sigma\sigma^{\prime}}\int\mathrm{d}\bm{r}\mathrm{d}\bm{r}^{\prime}\frac{e^{2}}{\left|\bm{r}-\bm{r}^{\prime}\right|}\hat{\psi}_{\sigma}^{\dagger}(\bm{r})\hat{\psi}_{\sigma^{\prime}}^{\dagger}(\bm{r}^{\prime})\hat{\psi}_{\sigma^{\prime}}(\bm{r}^{\prime})\hat{\psi}_{\sigma}(\bm{r})}}\\
\underset{\hat{V}_{\mathrm{ei}}}{\underbrace{-\sum_{i=1}^{N_{\mathrm{i}}}\sum_{\sigma}\int\mathrm{d}\bm{r}\frac{Z_{i}e^{2}}{\left|\bm{r}-\bm{R}_{i}\right|}\hat{\psi}_{\sigma}^{\dagger}(\bm{r})\hat{\psi}_{\sigma}(\bm{r})}}\underset{\hat{H}_{\mathrm{i}}}{\underbrace{-\sum_{i}\frac{\hbar^{2}}{2M_{i}}\nabla_{\bm{R}_{i}}^{2}+\frac{1}{2}\sum_{ij}\frac{Z_{i}Z_{j}e^{2}}{\left|\bm{R}_{i}-\bm{R}_{j}\right|}}}
\end{multline}
where the first two terms form the Hamiltonian of an electron subsystem,
expressed in the second quantized form, the third term is the interaction
between electrons and ions, and the last two terms form the Hamiltonian
of an ion subsystem. For ions, we use the first quantized form because
the exchange symmetry will be ignored in following considerations.
The partition function of the system is determined by $Z=\mathrm{Tr}e^{-\beta\hat{K}}$.

We apply the classical isomorphism~\citep{chandler1981} to the ion
degrees of freedom. This is to interpret $e^{-\beta\hat{K}}$ as a
time evolution operator in the interval $[0,\hbar\beta)$ of the imaginary
time $t\equiv-\mathrm{i}\tau$, divide the interval into $N_{\mathrm{b}}$-slices,
and insert the closure relation $\int\mathrm{d}\bm{R}\ket{\bm{R}}\bra{\bm{R}}=1$
between the slices:
\begin{equation}
\mathrm{Tr}e^{-\beta\hat{K}}=\mathrm{Tr}\prod_{a=0}^{N_{\mathrm{b}}-1}e^{-\Delta\tau\hat{K}/\hbar}=\mathrm{Tr}_{\mathrm{e}}\int\left[\prod_{a=0}^{N_{\mathrm{b}}-1}\mathrm{d}\bm{R}\left(\tau_{a}\right)\right]\prod_{a=0}^{N_{\mathrm{b}}-1}\Braket{\bm{R}\left(\tau_{a+1}\right)|e^{-\Delta\tau\hat{K}/\hbar}|\bm{R}\left(\tau_{a}\right)},
\end{equation}
where $\Delta\tau\equiv\tau_{a+1}-\tau_{a}=\hbar\beta/N_{\mathrm{b}}$,
$\mathrm{Tr}_{\mathrm{e}}$ denotes the trace over electron degrees
of freedom, and the trace over ion degrees of freedom is taken care
by the path integrals over $\bm{R}(\tau_{a})$ and the periodic boundary
condition $\bm{R}(\tau_{N_{\mathrm{b}}})=\bm{R}(\tau_{0})$.

We can then apply the standard approximation of the path-integral
formalism to evaluate the matrix elements of the evolution operator
in a small time-interval $\Delta\tau$~\citep{negele1988}, and obtain~\citep{chandler1981}:
\begin{align}
Z & =\lim_{N_{\mathrm{b}}\rightarrow\infty}\int\left[\prod_{a=1}^{N_{\mathrm{b}}-1}\left(\frac{mN_{\mathrm{b}}}{2\pi\hbar^{2}\beta}\right)^{3/2}\mathrm{d}\bm{R}\left(\tau_{a}\right)\right]\left\{ \mathrm{Tr}_{\mathrm{e}}\prod_{a=1}^{N_{\mathrm{b}}-1}e^{-\Delta\tau\left[\hat{K}_{\mathrm{e}}+\hat{V}_{\mathrm{ei}}\left(\bm{R}(\tau_{a})\right)\right]/\hbar}\right\} e^{^{-\beta H_{\mathrm{i}}^{\mathrm{C}}[\bm{R}(\tau)]}}\\
 & \equiv\int\mathrm{D}\left[\bm{R}(\tau)\right]\left\{ \mathrm{Tr}\hat{T}_{\tau}e^{-\frac{1}{\hbar}\int_{0}^{\hbar\beta}\mathrm{d}\tau\left[\hat{K}_{\mathrm{e}}+\hat{V}_{\mathrm{ei}}(\tau)\right]}\right\} e^{^{-\beta H_{\mathrm{i}}^{\mathrm{C}}[\bm{R}(\tau)]}},\label{eq:Zpath}\\
H_{\mathrm{i}}^{\mathrm{C}}[\bm{R}(\tau)] & \equiv\frac{mN_{\mathrm{b}}}{2\hbar^{2}\beta^{2}}\sum_{i=1}^{N_{\mathrm{i}}}\sum_{a=0}^{N_{\mathrm{b}}-1}\left|\boldsymbol{R}_{i}\left(\tau_{a+1}\right)-\boldsymbol{R}_{i}\left(\tau_{a}\right)\right|^{2}+\frac{1}{2N_{\mathrm{b}}}\sum_{a=1}^{N_{\mathrm{b}}-1}\sum_{ij}\frac{Z_{i}Z_{j}e^{2}}{\left|\bm{R}_{i}\left(\tau_{a}\right)-\bm{R}_{j}\left(\tau_{a}\right)\right|}.
\end{align}
We note that $\mathrm{Tr}$ in Eq.~(\ref{eq:Zpath}) stands for $\mathrm{Tr}_{\mathrm{e}}$
with the subscript dropped for brevity.

\end{widetext}

In the limit of $N_{\mathrm{b}}\rightarrow\infty$, Eq.~(\ref{eq:Zpath})
is an exact decomposition for the electron-ion coupled system except
that the exchange symmetry between ions is ignored. It decomposes
the system into a \emph{quantum} electron system subjected to an \emph{imaginary-time-dependent}
ionic field and a classical ensemble in which each ion is mapped into
a $\tau$-loop. 

In the opposite limit of $N_{\mathrm{b}}=1$, the decomposition becomes
the Born-Oppenheimer approximation, which is employed in classical
molecular dynamics. All information concerning the $\tau$-dependences
and therefore the imaginary-time dynamics will be lost in this limit.
Since EPC is intrinsically a dynamic process, it is essential to use
the PIMD instead of the classical molecular dynamics for extracting
its information. We emphasis that for determining equilibrium properties,
one only needs the information of the imaginary-time (as opposed to
the real-time) dynamics~\citep{fetter2003,mahan2000}, which is exactly
what a PIMD is simulated for.

A PIMD simulation basically samples a classical ensemble which is
governed by the effective Hamiltonian $H_{\mathrm{eff}}\left[\bm{R}(\tau)\right]=H_{\mathrm{i}}^{\mathrm{C}}\left[\bm{R}(\tau)\right]+\Omega_{\mathrm{ei}}\left[\bm{R}(\tau)\right]$
with $\Omega_{\mathrm{ei}}\equiv-(1/\beta)\ln Z_{\mathrm{ei}}$, where
$Z_{\mathrm{ei}}$ is the expression inside the curly bracket in Eq.~(\ref{eq:Zpath}).
It is necessary to use a finite $N_{\mathrm{b}}$ in the simulation.
As a result, each quantum ion is mapped into a ring-polymer with $N_{\mathrm{b}}$
beads. In this case, the information of the imaginary-time dynamics
is preserved in the dependences of various functions on the discretized
imaginary time or the beads. The discretization inevitably causes
the loss of information and introduces errors. In circumstances, one
has to find ways to control the errors. See Sec.~\ref{subsec:Density-response-function}
for such an example.

With the decomposition, the evaluation of an electron-related quantity
becomes a two-step process. For instance, to determine the single-particle
Green's function of electrons, we have:
\begin{align}
\bar{\mathcal{G}}(\tau,\tau^{\prime})\equiv & -\frac{1}{Z}\mathrm{Tr}\left[\hat{T}_{\tau}\hat{\psi}(\tau)\hat{\psi}^{\dagger}(\tau^{\prime})e^{-\frac{1}{\hbar}\int_{0}^{\hbar\beta}\mathrm{d}\tau\hat{K}}\right]\\
= & -\frac{1}{Z}\int\mathrm{D}\left[\bm{R}(\tau)\right]e^{^{-\beta\left(H_{\mathrm{i}}^{\mathrm{C}}+\Omega_{\mathrm{ei}}\right)}}\frac{1}{Z_{\mathrm{ei}}}\nonumber \\
 & \times\mathrm{Tr}\hat{T}_{\tau}\hat{\psi}(\tau)\hat{\psi}^{\dagger}(\tau^{\prime})e^{-\frac{1}{\hbar}\int_{0}^{\hbar\beta}\mathrm{d}\tau\left[\hat{K}_{\mathrm{e}}+\hat{V}_{\mathrm{ei}}(\tau)\right]}\\
\equiv & \left\langle \mathcal{G}\left[\bm{R}(\tau)\right]\left(\tau,\tau^{\prime}\right)\right\rangle _{\mathrm{C}},
\end{align}
where $\mathcal{G}\left[\bm{R}(\tau)\right]$ is defined in Eq.~(\ref{eq:Ges}),
and $\left\langle \dots\right\rangle _{\mathrm{C}}$ denotes the classical
ensemble average over ion trajectories.

\subsubsection{Effective Pairing Interaction\label{subsec:Effective-Pairing-Interaction}}

With the preparations, we are ready to establish the three equations
summarized in Sec.~\ref{subsec:Effective-Interaction-Mediated}.
From Eq.~(\ref{eq:Wchi}), we see that to determine the effective
pairing interaction $W$, one needs to first determine the anomalous
response function $\hat{\chi}$. By treating the electron-subsystem
as an effective non-interacting system, we can apply Wick's theorem,
and obtain: 
\begin{equation}
\chi_{11^{\prime}}=-\frac{1}{\hbar}\left\langle \mathcal{G}_{11^{\prime}}[\bm{R}(\tau)]\mathcal{G}_{\bar{1}\bar{1^{\prime}}}[\bm{R}(\tau)]\right\rangle _{\mathrm{C}}.
\end{equation}
The pair scattering amplitude appeared in Eq.~(\ref{eq:Gamma}) is
defined by the decomposition
\begin{equation}
\hat{\chi}=\hat{\chi}_{0}+\frac{1}{\hbar\beta}\hat{\chi}_{0}\hat{\Gamma}\hat{\chi}_{0}.\label{eq:chidecompose}
\end{equation}
It is easy to verify that $\Gamma_{11^{\prime}}\equiv-\beta\langle\mathcal{T}_{11^{\prime}}[\bm{R}(\tau)]\mathcal{T}_{\bar{1}\bar{1^{\prime}}}[\bm{R}(\tau)]\rangle_{\mathrm{C}}$
with $\hat{\mathcal{T}}\equiv\hbar\hat{\bar{\mathcal{G}}}^{-1}(\hat{\mathcal{G}}[\bm{R}(\tau)]-\hat{\bar{\mathcal{G}}})\hat{\bar{\mathcal{G}}}^{-1}$.
We thus obtain Eq.~(\ref{eq:Gamma}). It is also easy to verify that
$\hat{\mathcal{T}}$ does satisfy Eq.~(\ref{eq:TRtau}). Finally,
by applying Eq.~(\ref{eq:Wchi}), it is not difficult to verify Eq.~(\ref{eq:W}).
It concludes our proof.

We still need to address the effect of the Coulomb interaction between
electrons since the above derivation treats the system as if it is
non-interacting. The Coulomb interaction introduces a number of revisions
to our result and derivation: (i) when determining the Green's function
$\mathcal{G}[\bm{R}(\tau)]$, one needs to introduce a self-energy
functional $\Sigma_{\mathrm{c}}[\mathcal{G}]$ which accounts for
the effect of the Coulomb interaction~\citep{potthoff2003} (see
Sec.~\ref{subsec:Effective-action-theory}). In practical calculations
which employ the DFT, the Green's function could be interpreted as
the Kohn-Sham Green's function with respect to an effective ionic
field $V_{\mathrm{ei}}^{\mathrm{KS}}\left[\bm{R}(\tau)\right]$ which
includes both the bare ionic potential and the screening potential
induced by the self-consistent electron density~\citep{kotliar2006};
(ii) when determining the anomalous response function in the time-dependent
quantum ensemble, there will be many-body corrections corresponding
to Feynman diagrams with at least one Coulomb interaction line (See
Fig.~19 of Ref.~\citep{scalapino1969}). As argued in the conventional
EPC theory, these contributions could be absorbed into renormalization
constants~\citep{scalapino1969}; (iii) the Luttinger-Ward functional
will have a component $\Omega_{\mathrm{LW}}^{(\mathrm{c})}[\mathcal{G},\mathcal{F}]$
contributed by the Coulomb interaction. It gives rise to a contribution
to $W$ arisen from $\left.\delta^{2}\Omega_{\mathrm{LW}}^{(\mathrm{c})}/\delta\mathcal{F}_{1}^{\ast}\delta\mathcal{F}_{1^{\prime}}\right|_{\mathcal{G}\rightarrow\bar{\mathcal{G}},\mathcal{F}\rightarrow0}$.
Its effect could be captured by an empirical Coulomb pseudopotential
parameter $\mu^{\ast}$ introduced in the conventional EPC theory~\citep{scalapino1969}.

\subsubsection{Linearized Eliashberg Equations\label{subsec:Linearized-Eliashberg-equation}}

\paragraph{Stiffness}

To estimate $T_{\mathrm{c}}$, we determine when a system becomes
unstable toward forming Cooper pairs. This is to exam the stiffness
matrix of the system with respect to the variations of the anomalous
Green's function $\mathcal{F}$. Because of the stiffness theorem
Eq.~(\ref{eq:stiffness}), the stiffness matrix is proportional to
$-\chi^{-1}$. Therefore, the non-negative-definiteness of $\chi^{-1}$
indicates an instability toward forming Cooper pairs and the superconducting
state. By applying Eq.~(\ref{eq:Wchi}), we have $\hat{\chi}^{-1}=\hat{\chi}_{0}^{-1}-(\hbar\beta)^{-1}\hat{W}$.
Because $\hat{\chi}_{0}$ is negative-definite, the negative-definiteness
of $\hat{\chi}^{-1}$ is equivalent to the requirement that the eigen-equation
\begin{equation}
\left(I-\frac{1}{\hbar\beta}\hat{W}\hat{\chi}_{0}\right)\hat{\Delta}=\rho\hat{\chi}_{0}\hat{\Delta}
\end{equation}
has no positive eigenvalue $\rho$.

The equation can be simplified. We have $[\hat{\chi}_{0}]_{11^{\prime}}=-\hbar^{-1}\left|\bar{\mathcal{G}}_{1}\right|^{2}\delta_{11^{\prime}}$,
and
\begin{align}
\left|\bar{\mathcal{G}}_{1}\right|^{2} & \equiv\left|\bar{\mathcal{G}}(\omega_{n},\bm{k})\right|^{2}\approx\frac{\pi\hbar^{2}}{\left|\tilde{\omega}(n)\right|}\delta\left(\tilde{\epsilon}_{\bm{k}}-\mu\right),
\end{align}
where we define a renormalized electron dispersion $\tilde{\epsilon}_{\bm{k}}=\epsilon_{\bm{k}}+\mathrm{Re}\bar{\Sigma}(\omega_{n},\bm{k})$
by ignoring the weak $\omega_{n}$-dependence of $\mathrm{Re}\bar{\Sigma}$,
and 
\begin{equation}
\tilde{\omega}(n)\equiv\omega_{n}-\frac{1}{\hbar}\mathrm{Im}\bar{\Sigma}(\omega_{n},\bm{k}_{\mathrm{F}})\label{eq:omegatilden}
\end{equation}
for a wave-vector $\bm{k}_{\mathrm{F}}$ on the Fermi surface. The
approximation is possible because $\hbar\left|\tilde{\omega}(n)\right|$
is much smaller than the typical energy scale of electrons, i.e.,
the Fermi energy.

We then insert the approximated form of $\hat{\chi}_{0}$ into the
eigen-equation and note that the resulting equation is closed for
$\Delta_{1^{\prime}}\equiv\Delta(\omega_{n^{\prime}},\bm{k}_{\mathrm{F}}^{\prime})$
in the subspace of all wave-vectors on the Fermi surface. Because
the system is isotropic, we can seek for an eigenvector $\hat{\Delta}$
which does not depend on the direction of $\bm{k}_{\mathrm{F}}^{\prime}$.
Therefore, $\Delta_{1^{\prime}}=\Delta(\omega_{n^{\prime}})\equiv\Delta_{n^{\prime}}$.
The eigen-equation becomes:
\begin{multline}
\sum_{n^{\prime}}\left[-\sum_{\bm{k}^{\prime}}W_{11^{\prime}}\delta\left(\tilde{\epsilon}_{\bm{k}^{\prime}}-\mu\right)-\frac{\hbar\beta}{\pi}\left|\tilde{\omega}(n)\right|\delta_{nn^{\prime}}\right]\Delta_{n^{\prime}}\\
=\rho\Delta_{n^{\prime}}.
\end{multline}
The Coulomb pseudo-potential $\mu^{\ast}$ is then inserted by hand.
The resulting equation is exactly Eq.~(\ref{eq:Eliashberg1}).

\paragraph{Generalized optical theorem}

To close the equation, we still need to determine $\tilde{\omega}(n)$.
In the conventional Eliashberg theory, the self-energy is determined
by the effective interaction through a perturbative equation like
Eq.~(\ref{eq:imsigma}). In our non-perturbative treatment, however,
the self-energy is assumed to be known\emph{ a priori}. In principle,
$\bar{\Sigma}$ can be determined directly with a PIMD simulation.
However, it is infeasible in practice. This is because the accurate
determination of $\bar{\mathcal{G}}$ requires a high-resolution of
the imaginary time, i.e., a large $N_{\mathrm{b}}$ in the PIMD simulation.
Inaccuracy may introduce inconsistency because the two Eliashberg
equations, in their conventional forms, involve the same set of parameters
$\lambda(n)$.

Fortunately, we are able to establish a generalized optical theorem~\citep{vanoosten1985}
for the \emph{imaginary part} of the self-energy with a form identical
to Eq.~(\ref{eq:imsigma}). The derivation is detailed as follows.

By applying the Dyson equation
\begin{equation}
\left\{ \hat{\bar{\mathcal{G}}}^{-1}-\frac{\hat{\mathcal{V}}}{\hbar}\right\} \hat{\mathcal{G}}=I,\label{eq:DysonG-1}
\end{equation}
and the relation $\hat{\bar{\mathcal{G}}}=\langle\hat{\mathcal{G}}\rangle_{\mathrm{C}}$,
we have $\langle\hat{\mathcal{V}}\hat{\mathcal{G}}\rangle_{\mathrm{C}}=0$.
By inserting the definition of $\hat{\mathcal{V}}$ and the identity
$\hat{\mathcal{G}}=\hat{\bar{\mathcal{G}}}+\hbar^{-1}\hat{\bar{\mathcal{G}}}\hat{\mathcal{T}}\hat{\bar{\mathcal{G}}}$,
we obtain 
\begin{equation}
\hat{\bar{\Sigma}}=\langle\hat{V}_{\mathrm{ei}}\rangle_{\mathrm{C}}+\frac{1}{\hbar}\langle\hat{V}_{\mathrm{ei}}\hat{\bar{\mathcal{G}}}\hat{\mathcal{T}}\rangle_{\mathrm{C}}.
\end{equation}
We make further manipulations 
\begin{multline}
\langle\hat{V}_{\mathrm{ei}}\hat{\bar{\mathcal{G}}}\hat{\mathcal{T}}\rangle_{\mathrm{C}}=\left\langle \left(\hat{V}_{\mathrm{ei}}-\hat{\Sigma}^{\dagger}\right)\hat{\bar{\mathcal{G}}}\hat{\mathcal{T}}\right\rangle _{\mathrm{C}}=\left\langle \hat{\mathcal{V}}^{\dagger}\hat{\bar{\mathcal{G}}}\hat{\mathcal{T}}\right\rangle _{\mathrm{C}}\\
=\left\langle \hat{\mathcal{T}}^{\dagger}\hat{\bar{\mathcal{G}}}\hat{\mathcal{T}}-\frac{1}{\hbar}\hat{\mathcal{T}}^{\dagger}\hat{\bar{\mathcal{G}}}^{\dagger}(\hat{V}_{\mathrm{ei}}-\hat{\bar{\Sigma}}^{\dagger})\hat{\bar{\mathcal{G}}}\hat{\mathcal{T}}\right\rangle _{\mathrm{C}},
\end{multline}
where, in the first line, we make use of $\langle\hat{\mathcal{T}}\rangle_{\mathrm{C}}=0$,
and from the first line to the second line, we apply Eq.~(\ref{eq:TRtau})
to replace $\hat{\mathcal{V}}^{\dagger}$ with $\hat{\mathcal{V}}^{\dagger}=\hat{\mathcal{T}}^{\dagger}-\hbar^{-1}\hat{\mathcal{T}}^{\dagger}\hat{\bar{\mathcal{G}}}^{\dagger}\hat{\mathcal{V}}^{\dagger}$.
By noting that $\hat{V}_{\mathrm{ei}}$ is Hermitian and $\bar{\mathcal{G}}$
and $\bar{\Sigma}$ are diagonal in a liquid, we have
\begin{multline}
\mathrm{Im}\bar{\Sigma}_{1}=\frac{1}{\hbar}\left[\mathrm{Im}\left\langle \hat{V}_{\mathrm{ei}}\hat{\bar{\mathcal{G}}}\hat{\mathcal{T}}\right\rangle _{\mathrm{C}}\right]_{11}=\frac{1}{\hbar}\sum_{1^{\prime}}\\
\left\langle \mathcal{T}_{1^{\prime}1}^{\ast}\left(\mathrm{Im}\bar{\mathcal{G}}_{1^{\prime}}\right)\mathcal{T}_{1^{\prime}1}-\frac{1}{\hbar}\mathcal{T}_{1^{\prime}1}^{\ast}\bar{\mathcal{G}}_{1^{\prime}}^{\ast}\left(\mathrm{Im}\bar{\Sigma}_{1^{\prime}}\right)\bar{\mathcal{G}}_{1^{\prime}}\mathcal{T}_{1^{\prime}1}\right\rangle _{\mathrm{C}}\\
=-\frac{1}{\hbar\beta}\sum_{1^{\prime}}\left[\mathrm{Im}\bar{\mathcal{G}}_{1^{\prime}}\Gamma_{1^{\prime}1}-\frac{1}{\hbar}\mathrm{Im}\bar{\Sigma}_{1^{\prime}}\left|\bar{\mathcal{G}}_{1^{\prime}}\right|^{2}\Gamma_{1^{\prime}1}\right],
\end{multline}
where we make use of Eq.~(\ref{eq:Gamma}). In the matrix form, the
equality can be written as
\begin{equation}
\mathrm{Im}\hat{\bar{\Sigma}}=-\frac{1}{\hbar\beta}\left(\mathrm{Im}\hat{\bar{\mathcal{G}}}\right)\hat{\Gamma}-\frac{1}{\hbar\beta}\left(\mathrm{Im}\hat{\bar{\Sigma}}\right)\hat{\chi}_{0}\hat{\Gamma}.
\end{equation}
We then have
\begin{align}
\mathrm{Im}\hat{\bar{\Sigma}} & =-\frac{1}{\hbar\beta}\left(\mathrm{Im}\hat{\bar{\mathcal{G}}}\right)\left[\hat{\Gamma}\left(I+\frac{1}{\hbar\beta}\hat{\chi}_{0}\hat{\Gamma}\right)^{-1}\right]\\
 & =-\frac{1}{\hbar\beta}\left(\mathrm{Im}\hat{\bar{\mathcal{G}}}\right)\hat{W},
\end{align}
where we make use of the matrix form of Eq.~(\ref{eq:W}) $\hat{W}=\hat{\Gamma}-(\hbar\beta)^{-1}\hat{W}\hat{\chi}_{0}\hat{\Gamma}$.
The final form is exactly the matrix form of Eq.~(\ref{eq:imsigma}).
By inserting Eq.~(\ref{eq:imsigma}) into Eq.~(\ref{eq:omegatilden}),
we obtain Eq.~(\ref{eq:on}).

We note that there is no simple relation like Eq.~(\ref{eq:imsigma})
for $\mathrm{Re}\Sigma$. Fortunately, $\mathrm{Re}\Sigma$ is dominated
by $\langle\hat{V}_{\mathrm{ei}}\rangle_{\mathrm{C}}$, and the correction
due to EPC is usually small and negligible (see Fig.~\ref{fig:self-energy}).

\subsubsection{Reducing to the conventional EPC theory\label{subsec:Reduction-of-the}}

The conventional EPC theory deals with crystalline solids and assumes
that the vibration amplitudes of ions are small. In the lowest order,
the ions could be regarded to be fixed in their respective equilibrium
lattice positions $\left\{ \bm{R}_{i}^{0}\right\} $. As a result,
the self-energy $\bar{\Sigma}$ can be approximated as:
\begin{equation}
\bar{\Sigma}\approx V_{\mathrm{ei}}^{(0)}\equiv V_{\mathrm{ei}}\left(\left\{ \bm{R}_{i}^{0}\right\} \right).
\end{equation}
One expects that the vibrations of ions will introduce a correction
to the self-energy, i.e., the EPC correction to the self-energy. Since
the vibration amplitudes are small, the correction is expected to
be small.

One can then determine a set of Bloch wave-functions $\varphi_{a\bm{k}}\equiv V^{-1/2}\exp(\mathrm{i}\bm{k}\cdot\bm{r})u_{a\bm{k}}(\bm{r})$
by solving the Schr\"{o}dinger equation in the presence of $V_{\mathrm{ei}}^{(0)}$,
where $\bm{k}$ is a quasi-wave-vector and $a$ is a band index. The
average Green's function will be approximately diagonal in the basis:
\begin{equation}
\bar{\mathcal{G}}_{11^{\prime}}=\bar{\mathcal{G}}_{1}\delta_{11^{\prime}}+\Delta\bar{\mathcal{G}}_{11^{\prime}},
\end{equation}
where the indices $1$ and $1^{\prime}$ correspond to the combinations
of $(\omega_{n},\bm{k},a)$, and $\Delta\bar{\mathcal{G}}_{11^{\prime}}$
denotes a small correction due to the vibrations of ions. By inspecting
Eqs.~(\ref{eq:TRtau}, \ref{eq:W}, \ref{eq:imsigma}), we find that
the correction $\Delta\bar{\mathcal{G}}_{11^{\prime}}$ can be ignored
since in these equations the Green's function is always multiplied
by small quantities like $\mathcal{V}$ and $W$.

The scattering potential can then be approximated as:
\begin{align}
\hat{\mathcal{V}} & \equiv\hat{V}_{\mathrm{ei}}-\hat{\bar{\Sigma}}\\
 & \approx\hat{V}_{\mathrm{ei}}\left(\left\{ \bm{R}_{i}\right\} \right)-\hat{V}_{\mathrm{ei}}\left(\left\{ \bm{R}_{i}^{0}\right\} \right)\\
 & \approx\sum_{i\alpha\kappa}\left.\frac{\partial\hat{V}_{\mathrm{ei}}}{\partial R_{i\alpha\kappa}}\right|_{\left\{ \bm{R}_{i}^{0}\right\} }u_{i\alpha\kappa},
\end{align}
where $i$, $\alpha$, $\kappa$ are indices of unit cells, axis directions
and sub-lattices, respectively, and $u_{i\alpha\kappa}\equiv R_{i\alpha\kappa}-R_{i\alpha\kappa}^{0}$
is the displacement of an ion. We know from the conventional EPC theory
that the correction to $\bar{\Sigma}$ due to ion vibrations is proportional
to $\left|\mathcal{V}\right|^{2}$, and is thus negligible.

The displacements of ions can be expressed in terms of phonon annihilation
and creation operators $\hat{a}_{\bm{q}\nu}$, $\hat{a}_{\bm{q}\nu}^{\dagger}$.
The scattering potential can then be written as (see Eq.~(32) of
Ref.~\citep{giustino2017}):
\begin{equation}
\hat{\mathcal{V}}=\frac{1}{\sqrt{N_{\mathrm{i}}}}\sum_{\bm{q}\nu}\Delta_{\bm{q}\nu}\hat{V}_{\mathrm{ei}}\left(\hat{a}_{\bm{q}\nu}+\hat{a}_{-\bm{q}\nu}^{\dagger}\right),
\end{equation}
where $\Delta_{\bm{q}\nu}\hat{V}_{\mathrm{ei}}$ is defined in Ref.~\citep{giustino2017}
(as $\Delta_{\bm{q}\nu}V^{\mathrm{KS}}$).

Since $\hat{\mathcal{V}}$ is a small quantity, we can apply the Born
approximation to Eq.~(\ref{eq:TRtau}) and obtain $\hat{\mathcal{T}}\approx\text{\ensuremath{\hat{\mathcal{V}}}}$.
The matrix elements of $\hat{\mathcal{T}}$ with respect to the basis
function $\varphi_{\omega_{n}a\bm{k}}(\bm{r}\tau)=(\hbar\beta V)^{-1/2}\exp\left(-\mathrm{i}\omega_{n}\tau+\mathrm{i}\bm{k}\cdot\bm{r}\right)u_{a\bm{k}}(\bm{r})$
are:
\begin{multline}
\mathcal{T}_{11^{\prime}}=\frac{1}{\sqrt{V}}g_{aa^{\prime}\nu}(\bm{k}^{\prime},\bm{q})\delta_{\bm{k},\bm{k}^{\prime}+\bm{q}}\\
\times\frac{1}{\hbar\beta}\int_{0}^{\hbar\beta}\mathrm{d}\tau\left[\hat{a}_{\bm{q}\nu}(\tau)+\hat{a}_{-\bm{q}\nu}^{\dagger}(\tau)\right]e^{-\mathrm{i}(\omega_{n}-\omega_{n^{\prime}})\tau}
\end{multline}
with the electron-phonon matrix element $g_{aa^{\prime}\nu}(\bm{k}^{\prime},\bm{q})$
defined in Eq.~(38) of Ref.~\citep{giustino2017}.

By applying Eq.~(\ref{eq:Gamma}), and noting that the path-integral
average $\left\langle \dots\right\rangle _{\mathrm{C}}$ is equivalent
to a time-ordered average of operators~\citep{negele1988}, we obtain:
\begin{equation}
\Gamma_{11^{\prime}}=\frac{1}{V}\delta_{\bm{k},\bm{k}^{\prime}+\bm{q}}\left|g_{aa^{\prime}\nu}(\bm{k}^{\prime},\bm{q})\right|^{2}\mathcal{D}_{\nu}(\bm{q},\omega_{n}),\label{eq:Gamma11}
\end{equation}
where the phonon Green's function is
\begin{equation}
\mathcal{D}(\bm{q},\omega_{n})=-\frac{1}{\hbar}\int_{0}^{\hbar\beta}\mathrm{d}\tau\left\langle \hat{T}_{\tau}\hat{A}_{\bm{q}\nu}(\tau)\hat{A}_{\bm{q}\nu}^{\dagger}(0)\right\rangle e^{\mathrm{i}\omega_{n}\tau},
\end{equation}
with $\hat{A}_{\bm{q}\nu}\equiv\hat{a}_{\bm{q}\nu}(\tau)+\hat{a}_{-\bm{q}\nu}^{\dagger}(\tau)$.

We then apply the Born-approximation to the BS equation (\ref{eq:W}),
and have $\hat{W}\approx\hat{\Gamma}$, i.e., Eq.~(\ref{eq:Gamma11})
is the effective interaction induced by EPC. The result should be
compared with its counterpart in the conventional EPC theory, see,
for instance, Eq.~(7.276) of Ref.~\citep{mahan2000}, in which the
electron-phonon matrix element is denoted as $M_{\lambda}(\bm{q})$.
It is easy to see that the two are equivalent.

\section{Numerical Implementation for Metallic Hydrogen\label{sec:Numerical-Implementation-for}}

Based on the formalism Eqs.~(\ref{eq:Gamma}--\ref{eq:lambda}),
we can develop a scheme for estimating $T_{\mathrm{c}}$. For samples
of ion trajectories from a PIMD simulation~\citep{chen2013}, $T$-matrices
are determined by solving Eq.~(\ref{eq:TRtau}). The pair scattering
amplitude is determined from the fluctuation of the $T$-matrices
by applying Eq.~(\ref{eq:Gamma}). The effective pairing interaction
is obtained from the scattering amplitude by solving Eq.~(\ref{eq:W}).
The interaction parameters $\lambda(n)$ are evaluated by using Eq.~(\ref{eq:lambda}).
The linearized Eliashberg equations~(\ref{eq:Eliashberg1}--\ref{eq:on})
are then solved, and the maximal eigenvalue $\rho_{\mathrm{m}}$ of
the equations are determined. With $\rho_{\mathrm{m}}$, we can determine
whether the temperature of the PIMD simulation is below ($\rho_{\mathrm{m}}>0$)
or above ($\rho_{\mathrm{m}}<0$) $T_{\mathrm{c}}$~\citep{rainer1974,allen1975}.
By varying the PIMD simulation temperature, $T_{\mathrm{c}}$ can
be estimated from the condition $\rho_{\mathrm{m}}=0$. The procedure
is detailed in Sec.~\ref{subsec:Numerical-implementation}.

To make the scheme practical for real calculations, we adopt the quasi-static
approximation. This is to treat the scattering potential $\hat{\mathcal{V}}(\tau)$
as a static potential, and solve Eq.~(\ref{eq:TRtau}) to obtain
a $\tau$-dependent $T$-matrix $\hat{\mathcal{T}}_{N_{\mathrm{s}}}(\tau)$
in the elastic limit by setting the frequency of $\bar{\mathcal{G}}$
to $\omega_{N_{\mathrm{s}}}\equiv(2N_{\mathrm{s}}+1)\pi/\hbar\beta$,
where $N_{\mathrm{s}}$ is a large integer satisfying $\omega_{\mathrm{ph}}\ll\omega_{N_{\mathrm{s}}}\ll\epsilon_{\mathrm{F}}/\hbar$
with $\omega_{\mathrm{ph}}$ being the scale of phonon frequencies
and $\epsilon_{\mathrm{F}}$ the Fermi energy of electrons. The $T$-matrix
is then approximated as $\hat{\mathcal{T}}(\omega_{N_{\mathrm{s}}}+\nu_{m},\omega_{N_{\mathrm{s}}})\approx(1/\hbar\beta)\int_{0}^{\hbar\beta}\mathrm{d}\tau\hat{\mathcal{T}}_{N_{\mathrm{s}}}(\tau)e^{i\nu_{m}\tau}$
for $\nu_{m}\equiv2m\pi/\hbar\beta$, $m\in Z$. We can show that
the quasi-static approximation becomes exact in the limit of $\omega_{N_{\mathrm{s}}}\gg\omega_{\mathrm{ph}}$.
With the approximation, we can determine effective pairing interaction
matrix elements $\hat{W}(\omega_{N_{\mathrm{s}}}+\nu_{m},\omega_{N_{\mathrm{s}}})$.
Physically, one expects that $\hat{W}(\omega_{n}+\nu_{m},\omega_{n})$
is close to $\hat{W}(\omega_{N_{\mathrm{s}}}+\nu_{m},\omega_{N_{\mathrm{s}}})$
as long as $|\omega_{n}-\omega_{N_{\mathrm{s}}}|\ll\epsilon_{\mathrm{F}}/\hbar$.
As a result, the effective pairing interaction can be determined by
assuming $\hat{W}(\omega_{n}+\nu_{m},\omega_{n})\approx\hat{W}(\omega_{N_{\mathrm{s}}}+\nu_{m},\omega_{N_{\mathrm{s}}})$.
See Sec.~\ref{subsec:Quasi-static-approximation} for details.

For metallic hydrogen, we use the linear screening approximation for
calculating the effective ionic potential for a given ionic configuration:
$V_{\mathrm{ei}}(\bm{q})\approx v_{\mathrm{ei}}(\boldsymbol{q})\rho_{\mathrm{i}}(\bm{q})/\epsilon_{\mathrm{et}}(\boldsymbol{q})$,
where $v_{\mathrm{ei}}(\bm{q})$ is the Coulomb interaction between
an electron and an ion, $\rho_{\mathrm{i}}(\bm{q})\equiv\sum_{i}\exp(-\mathrm{i}\bm{q}\cdot\bm{R}_{i})$,
and $\epsilon_{\mathrm{et}}(\boldsymbol{q})$ is the static electron-test
charge dielectric function~\citep{giuliani2005} with Ichimaru-Utsumi's
local field correction factor~\citep{ichimaru1981}. Compared to
the self-consistent Kohn-Sham potential determined by the DFT, the
approximation is only a few percent off, as shown in the inset of
Fig.~\ref{fig:W}. The precision is sufficient for implementing and
testing a new approach.

\subsection{Numerical implementation\label{subsec:Numerical-implementation}}

We implement our scheme as an add-on to existing PIMD simulations.
We first run a PIMD simulation which outputs samples of ion trajectories.
Each sample of the ion trajectories contains a set of coordinates
$\{\bm{R}_{i}(\tau_{a}),i=1\dots N_{\mathrm{i}},a=1\dots N_{\mathrm{b}}\}$,
where $N_{\mathrm{i}}$ is the total number of ions and $N_{\mathrm{b}}$
is the number of beads discretizing the imaginary time~\citep{chandler1981}.
The output then serves as the input of a program implementing our
scheme.

Our PIMD simulations are performed as in Ref.~\citep{chen2013} using
the Vienna \emph{ab initio} Simulation Package (VASP) code~\citep{vasp1,vasp2},
along with an implementation of the PIMD method used in Ref.~\citep{Feng2015}.
For metallic hydrogen, the implementation yields quantitatively the
same results as the one used in Ref.~\citep{chen2013} but with improved
sampling efficiency. The electronic structure was described ``on-the-fly''
using DFT. Projector augmented wave (PAW) potentials along with a
500~eV energy cutoff were employed for the expansion of the electronic
wave functions~\citep{paw1,paw2}. The Perdew-Burke-Ernzerhof (PBE)
functional was used to describe the electronic exchange-correlation
interaction~\citep{PBE}. The liquid state was modeled with a supercell
containing 200 atoms and a Monkhorst-Pack $\bm{k}$-point mesh of
spacing no larger than $2\pi\times0.05\text{Å}$ were used to sample
the Brillouin zone. The \emph{ab initio} PIMD simulations were performed
at $350\,\mathrm{K}$ and $\mathrm{450\,\mathrm{K}}$ with pressures
ranging from $0.5\,\mathrm{TPa}$ to $1.5\,\mathrm{TPa}$. The Andersen
thermostat was chosen to control the temperature of the canonical
(NVT) ensemble~\citep{Andersen1980}, in which the ionic velocities
were periodically randomized with respect to the Maxwellian distribution
every 25~fs. No less than 1.5~ps simulation length with $N_{\mathrm{b}}=24$
were used to evaluate the quantum fluctuation.

\begin{figure*}
\includegraphics[width=0.75\textwidth]{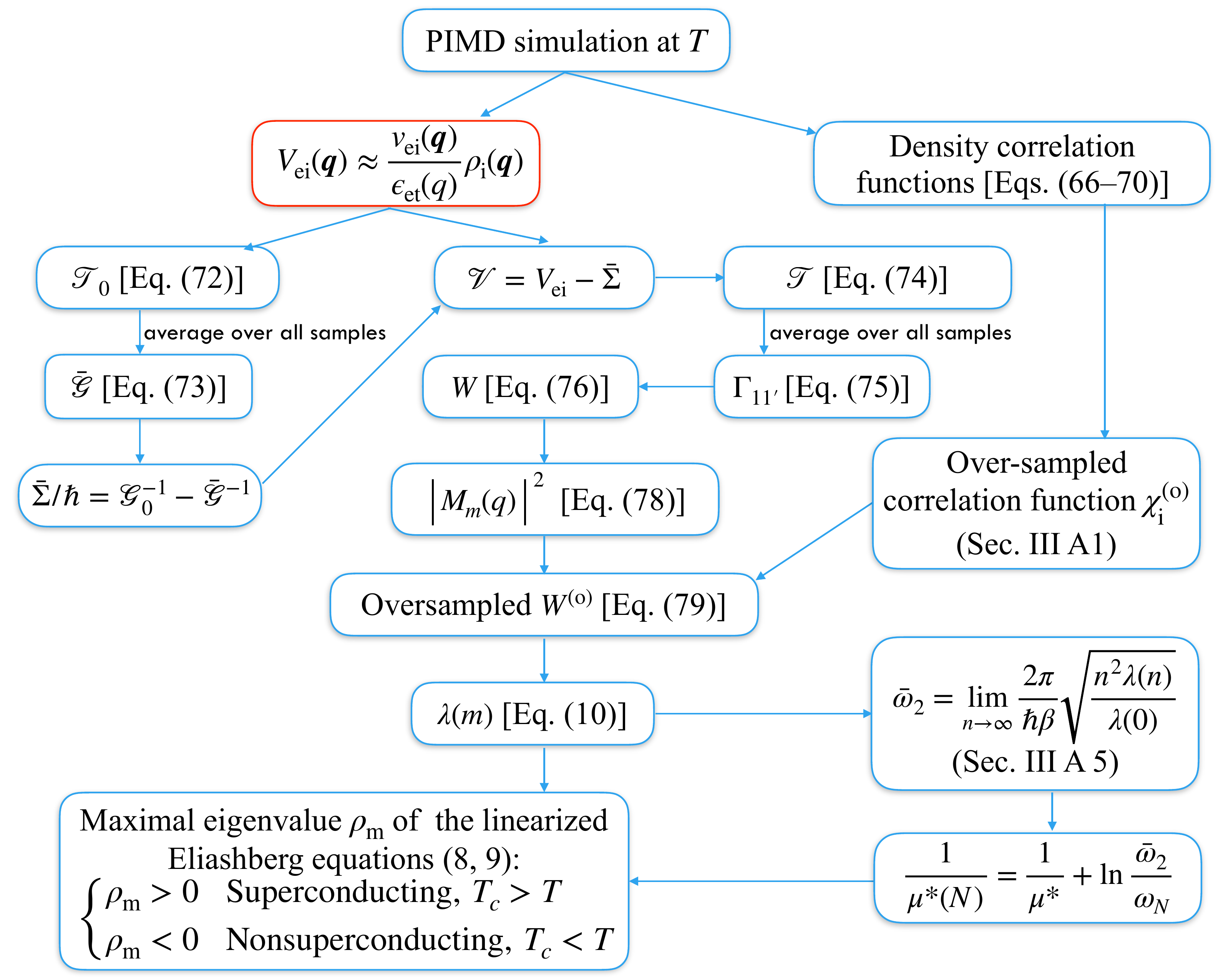}

\caption{\label{fig:Flowchart}Flowchart of the program for analyzing PIMD
outputs. $N$ denotes the cutoff of maximal $n$ when solving Eq.~(\ref{eq:Eliashberg1}).
The linear screening approximation, which should be replaced in a
full implementation, is indicated by the red box.}
\end{figure*}

Our program for analyzing PIMD outputs is implemented in MATLAB. Figure
\ref{fig:Flowchart} shows the flowchart of the program. The program
determines whether a PIMD simulation temperature is below or above
$T_{\mathrm{c}}$. To estimate $T_{\mathrm{c}}$, one needs to run
PIMD simulations at (at least) two different temperatures between
which the maximal eigenvalue $\rho_{\mathrm{m}}$ of the linearized
Eliashberg equations (\ref{eq:Eliashberg1}, \ref{eq:on}) changes
sign. $T_{\mathrm{c}}$ is estimated by a linear interpolation from
the two temperatures~\footnote{The source codes of the program can be downloaded from https://github.com/junrenshi/MetallicHydrogen.}.

In the following, we demonstrate our analyses by using the case of
$P=0.7\,\mathrm{TPa}$ and $T=350\,\mathrm{K}$ as an example.

\subsubsection{Density correlation function\label{subsec:Density-response-function}}

In a PIMD, the density correlation function can be decomposed into
two parts, including the self-correction function $\omega(\bm{q},\nu_{m})$
and the direct correlation function $h(\bm{q},\nu_{m})$:
\begin{equation}
\chi_{\mathrm{i}}(\bm{q},\nu_{m})=-\beta\rho_{0}\left[h(\bm{q},\nu_{m})+\omega(\bm{q},\nu_{m})\right],\label{eq:chii}
\end{equation}
where $\rho_{0}$ is the density of ions, and the definitions of the
various correlation functions can be found in Ref.~\citep{chandler1981}.
The self-correlation function is where the quantum effect is manifested.

To numerically evaluate the correlation functions, we first determine
for each sample of the ion trajectories:
\begin{align}
\tilde{\rho}_{i}(\bm{q},\nu_{m}) & =\frac{1}{N_{\mathrm{b}}}\sum_{a=1}^{N_{\mathrm{b}}}e^{-i\bm{q}\cdot\bm{R}_{i}(\tau_{a})+i\nu_{m}\tau_{a}},\\
\rho_{\mathrm{i}}(\bm{q},\nu_{m}) & =\sum_{i=1}^{N_{\mathrm{i}}}\tilde{\rho}_{i}(\bm{q},\nu_{m}).
\end{align}
The density correlation function $\chi_{\mathrm{i}}(\bm{q},\nu_{m})$
and the self-correction function $\omega(\bm{q},\nu_{m})$ can then
be determined by:
\begin{align}
\chi_{\mathrm{i}}(\bm{q},\nu_{m}) & =-\frac{\beta\rho_{0}}{N_{\mathrm{i}}}\left\langle \left|\rho_{\mathrm{i}}(\bm{q},\nu_{m})-\left\langle \rho_{\mathrm{i}}(\bm{q},\nu_{m})\right\rangle _{\mathrm{C}}\right|^{2}\right\rangle _{\mathrm{C}},\\
\omega(\bm{q},\nu_{m}) & =\left\langle \frac{1}{N_{\mathrm{i}}}\sum_{i=1}^{N_{\mathrm{i}}}\left|\tilde{\rho}_{i}(\bm{q},\nu_{m})\right|^{2}\right\rangle _{\mathrm{C}},
\end{align}
The direct correlation function $h(\bm{q},\nu_{m})$ can be determined
by applying the identity Eq.~(\ref{eq:chii}).

The finite number of the beads introduces discretization errors in
the determination of the correlation functions. It is the self-correlation
function which is prone to the discretization errors. This can be
seen in the self-correlation function of a free system~\citep{nichols1987}:
\begin{equation}
\omega_{0}(\bm{q},\tau)=\exp\left[-\frac{1}{2}(q\lambda_{e})^{2}\frac{\tau}{\hbar\beta}\left(1-\frac{\tau}{\hbar\beta}\right)\right],
\end{equation}
which becomes a sharp function of $\tau$ when $q$ is large, and
cannot be accurately sampled by a small number of $N_{\mathrm{b}}$
beads.

To solve the issue, we apply an over-sampling approach. A simulation
of $N_{\mathrm{b}}$ beads will give rise to a discrete set of values
of $\{\omega(\bm{q},\tau_{a}),\,a=1\dots N_{\mathrm{b}}\}$. We exploit
the property that $\ln\omega(\bm{q},\tau)$ is a smooth function of
$\tau$, and over-sample it by interpolating from its discrete set
of values. The resulting $\omega(\bm{q},\tau)$ can then be Fourier
transformed to obtain an over-sampled self-correlation function $\omega^{(\mathrm{o})}(\bm{q},\nu_{m})$.
By replacing $\omega(\bm{q},\nu_{m})$ with $\omega^{(\mathrm{o})}(\bm{q},\nu_{m})$
in Eq.~(\ref{eq:chii}), we can get an over-sampled density correlation
function $\chi_{\mathrm{i}}^{(\mathrm{o})}(\bm{q},\nu_{m})$, which
will be used in determining the interaction parameters (see Sec.~\ref{subsec:Interaction-parameters-and}).

\begin{figure}
\includegraphics[width=1\columnwidth]{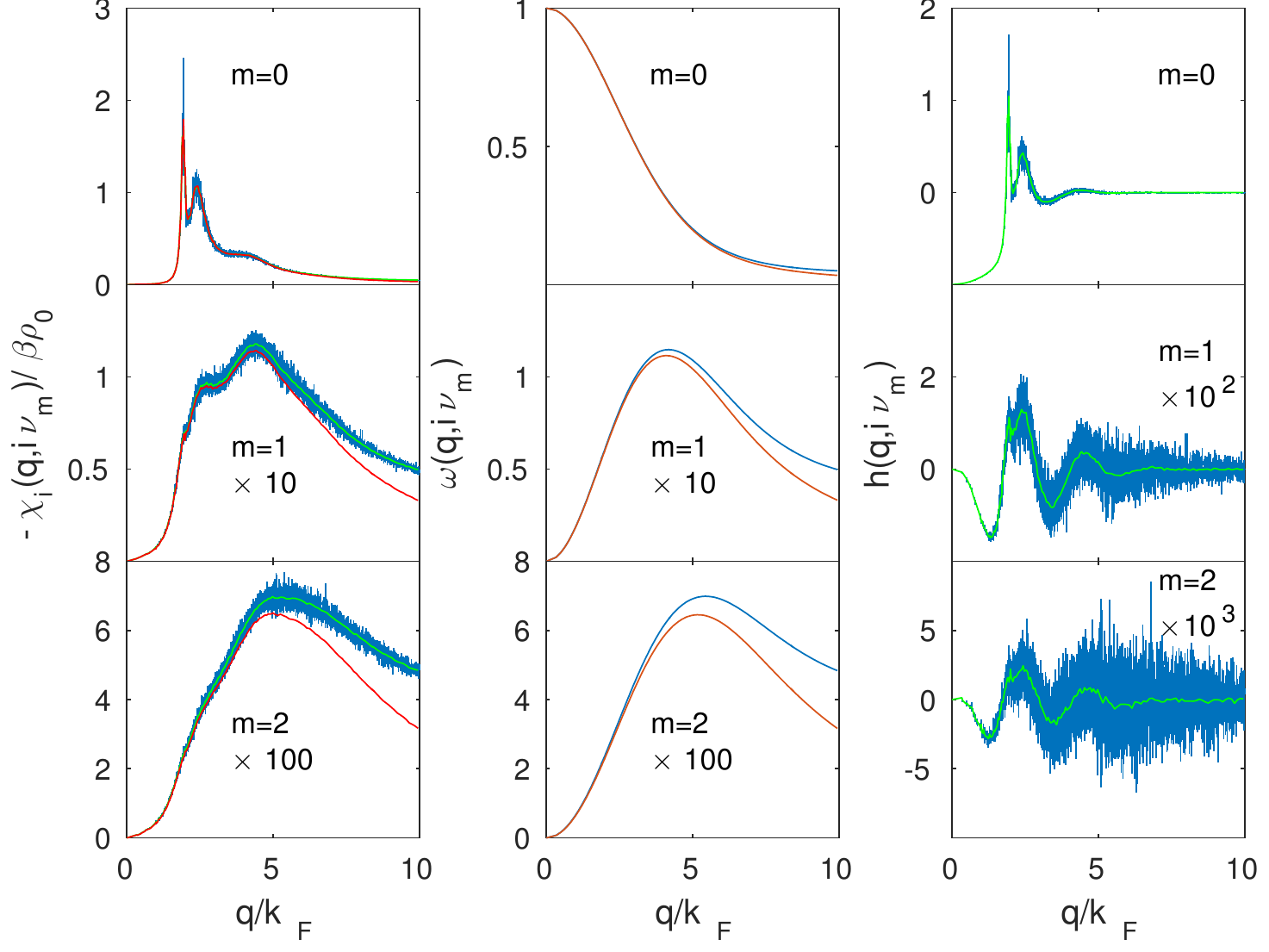}

\caption{\label{fig:Density-response-function}Various correlation functions
of ions. For the density correlation function and the self correlation
function, both the original one (blue) and over-sampled one (red)
are shown. The data uncertainties are estimated from the fluctuation
of values for the same $q$ but different $\bm{q}$'s, and indicated
by vertical lines extended from/to $\pm1$ standard deviation. The
over-sampling is with an increased number of beads $N_{\mathrm{b}}^{\prime}=16N_{\mathrm{b}}$.}
\end{figure}

The correlation functions of ions are shown in Fig.~\ref{fig:Density-response-function}.

\subsubsection{Lippmann-Schwinger equation}

\begin{figure}
\includegraphics[width=1\columnwidth]{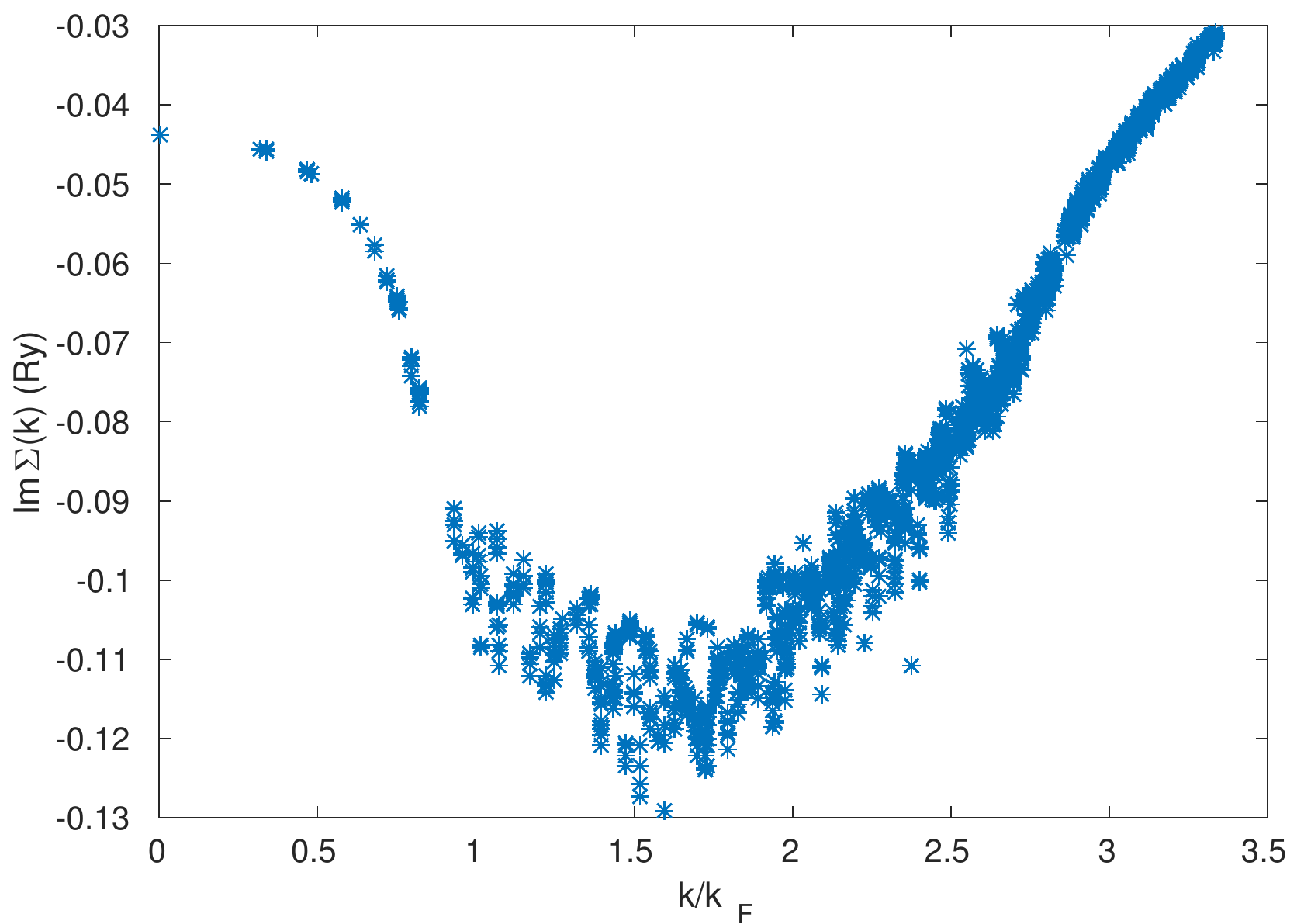}

\includegraphics[width=1\columnwidth]{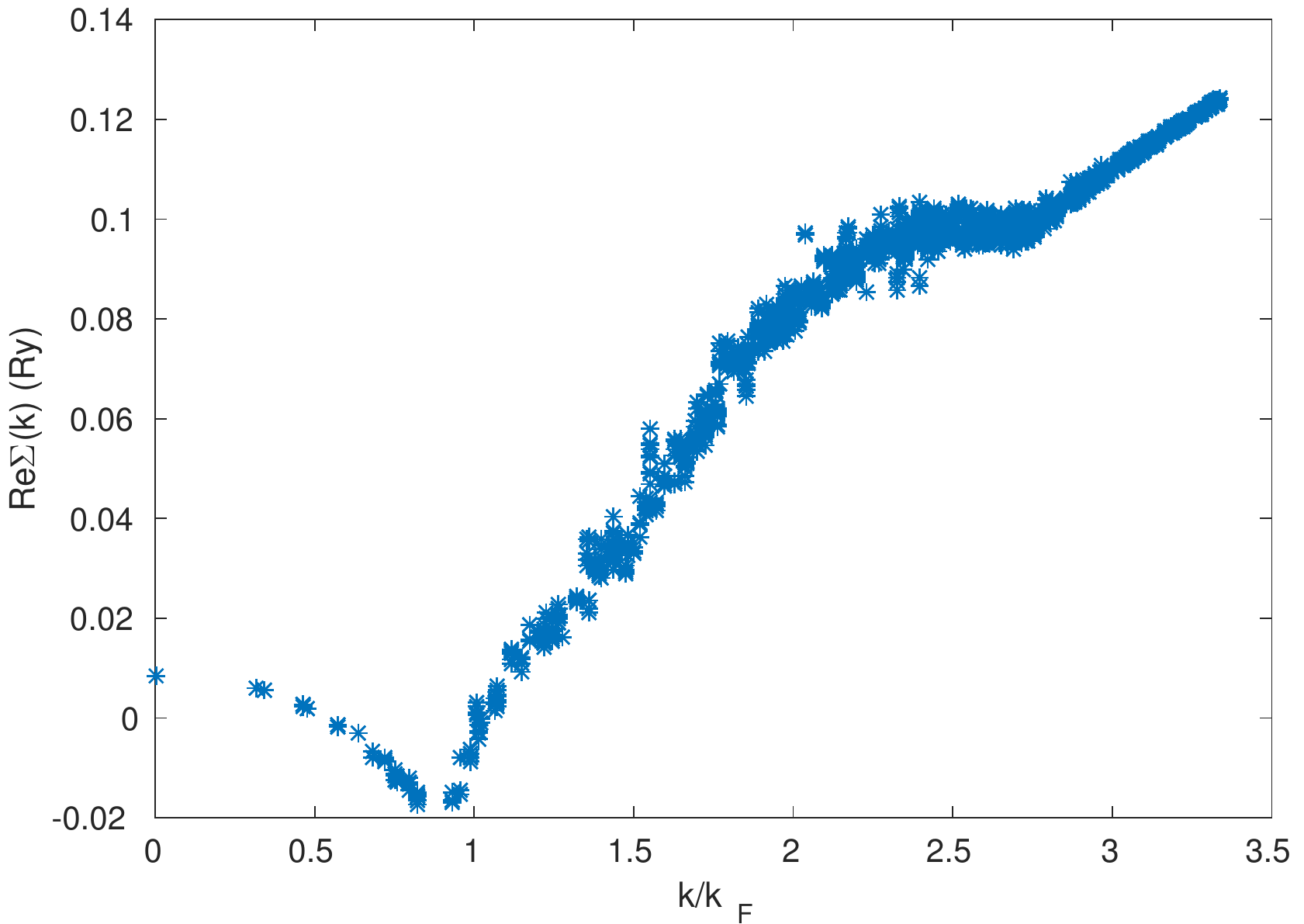}

\caption{\label{fig:self-energy}The imaginary and real parts of the self energy
$\bar{\Sigma}$. It is evaluated in the quasi-static limit with $N_{s}=16$.}
\end{figure}

We solve the Lippmann-Schwinger equation in the plane wave basis by
imposing an energy cutoff of $30\,\mathrm{Ry}$. With the quasi-static
approximation (see Sec.~\ref{subsec:Quasi-static-approximation}),
the $T$-matrix with respect to the \emph{vacuum} can be obtained
by:
\begin{equation}
\hat{\mathcal{T}}_{0}(\tau)=\left[I-\hat{V}_{\mathrm{ei}}(\tau)\hat{\mathcal{G}}_{0}\left(\omega_{N_{\mathrm{s}}}\right)\right]^{-1}\hat{V}_{\mathrm{ei}}(\tau),
\end{equation}
where $[\hat{\mathcal{G}}_{0}(\omega_{N_{\mathrm{s}}})]_{\bm{k}_{1},\bm{k}_{2}}=(i\omega_{N_{\mathrm{s}}}+\mu/\hbar-\hbar k_{1}^{2}/2m_{e})^{-1}\delta_{\bm{k}_{1},\bm{k}_{2}}$,
and $\hat{V}_{\mathrm{ei}}(\tau)$ is the effective ionic potential
at the imaginary time $\tau$. The average Green's function can be
obtained by applying the identity:
\begin{multline}
\hat{\bar{\mathcal{G}}}(\omega_{N_{\mathrm{s}}})=\mathcal{\hat{G}}_{0}(\omega_{N_{\mathrm{s}}})\\
+\frac{1}{\hbar}\mathcal{\hat{G}}_{0}(\omega_{N_{\mathrm{s}}})\left\langle \frac{1}{N_{\mathrm{b}}}\sum_{a}\hat{\mathcal{T}}_{0}(\tau_{a})\right\rangle _{\mathrm{C}}\hat{\mathcal{G}}_{0}(\omega_{N_{\mathrm{s}}}).
\end{multline}
The self-energy $\bar{\Sigma}(\omega_{N_{\mathrm{s}}})$ can be determined
from $\bar{\mathcal{G}}(\omega_{N_{\mathrm{s}}})$, and is shown in
Fig.~\ref{fig:self-energy}. A linear fitting to the real part of
the self-energy for $|k-k_{F}|<0.1k_{F}$ shows that the renormalization
to the Fermi velocity is only $\sim1.8\%$. It is ignored in our calculation
of the interaction parameters.

The $T$-matrix with respect to the effective medium can then be determined
by:
\begin{equation}
\hat{\mathcal{T}}(\tau)=\left(I-\hat{\mathcal{V}}(\tau)\hat{\bar{\mathcal{G}}}\right)^{-1}\hat{\mathcal{V}}(\tau),
\end{equation}
with $\hat{\mathcal{V}}(\tau)\equiv\hat{V}_{\mathrm{ei}}(\tau)-\hat{\bar{\Sigma}}$.
By applying a Fourier transform {[}see Eq.~(\ref{eq:TFourier}){]},
we obtain $\mathcal{T}_{\bm{k}_{1}\bm{k}_{2}}(\nu_{m})$. The scattering
amplitude is determined by:
\begin{equation}
\Gamma_{\bm{k}_{1}\bm{k}_{2}}(\nu_{m})=-\beta\left\langle \left|\mathcal{T}_{\bm{k}_{1}\bm{k}_{2}}(\nu_{m})\right|^{2}\right\rangle _{\mathrm{C}},
\end{equation}
and is shown in Fig.~\ref{fig:Scattering-amplitude-and}.

\subsubsection{Bethe-Salpeter equation\label{subsec:Bethe-Salpeter-equation}}

The effective pairing interaction can be obtained by solving the BS
equation in the quasi-static limit. The effective pairing interaction
at $\tau$ is determined by:
\begin{equation}
\hat{W}(\tau)=\left[I+\frac{1}{\hbar\beta}\hat{\Gamma}(\tau)\hat{\chi}_{0}\left(\omega_{N_{\mathrm{s}}}\right)\right]^{-1}\hat{\Gamma}(\tau),
\end{equation}
where $\hat{\Gamma}(\tau)$ is the Fourier transform of $\hat{\Gamma}(\nu_{m})$,
and $\hat{\chi}_{0}$ is a diagonal matrix with elements $-\hbar^{-1}|\bar{\mathcal{G}}(\omega_{N_{\mathrm{s}}},\bm{k})|^{2}$.
The effective interaction $W_{\bm{k}_{1}\bm{k}_{2}}(\nu_{m})$ can
then be obtained by a Fourier transform:
\begin{equation}
\hat{W}\left(\nu_{m}\right)=\frac{1}{\hbar\beta}\int_{0}^{\hbar\beta}\mathrm{d}\tau\hat{W}(\tau)e^{\mathrm{i}\nu_{m}\tau},
\end{equation}
 where $\nu_{m}\equiv2\pi m/\hbar\beta$ is a Bosonic Matsubara frequency.

Because $|\bar{\mathcal{G}}|^{2}$ is a function sharply peaks at
the Fermi surface, the equation can be solved in a truncated space
span by bases with wave vectors close to the Fermi surface. In our
calculation, we set a truncating condition $0.5k_{F}<|\bm{k}|<1.5k_{F}$.
We numerically confirm that varying the truncating condition does
not affect results.

\begin{figure}
\includegraphics[width=1\columnwidth]{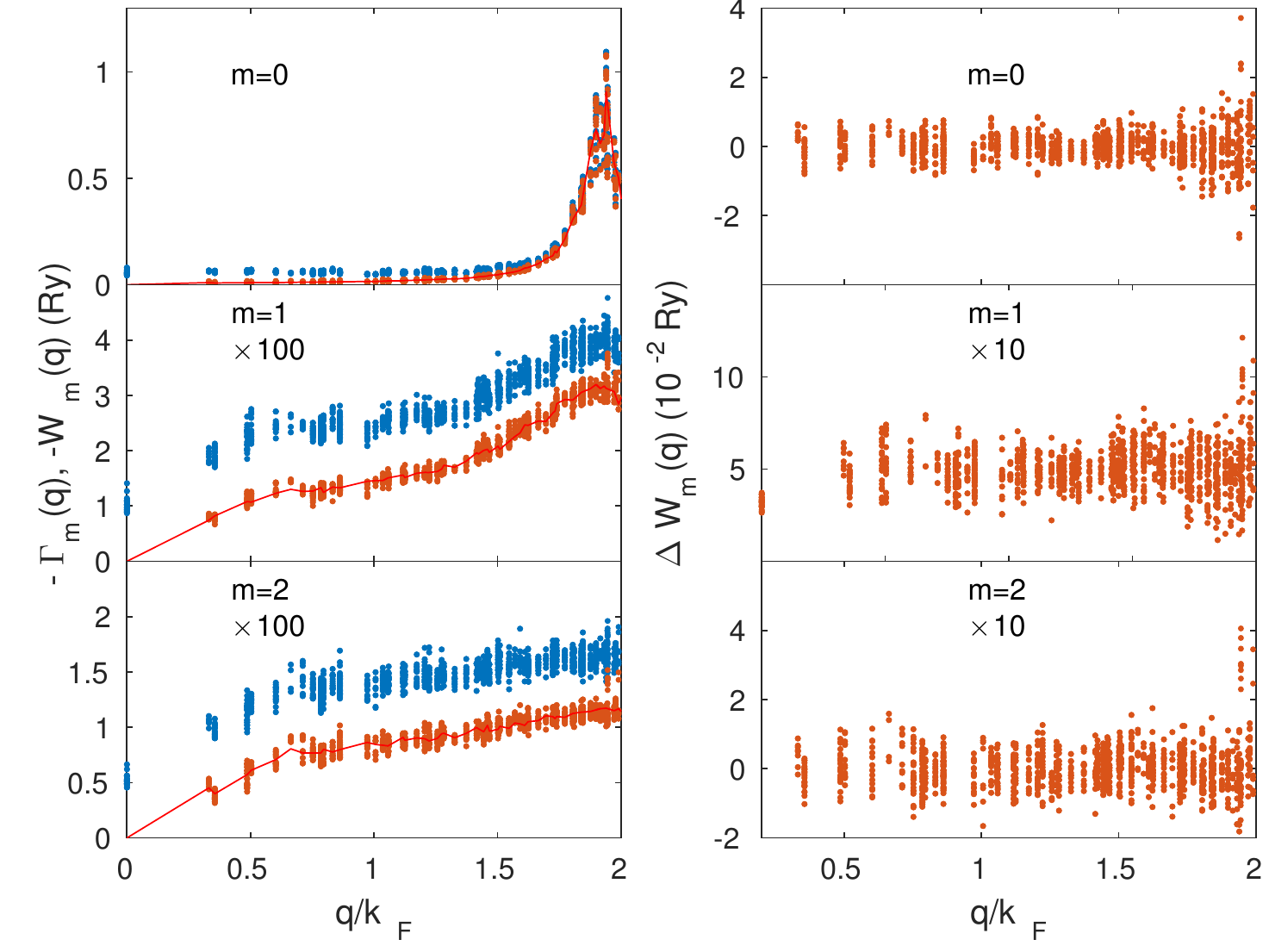}

\caption{\label{fig:Scattering-amplitude-and}Scattering amplitude and the
effective pairing interaction. Left: the values of $\Gamma_{\bm{k}_{1}\bm{k}_{2}}(\nu_{m})$
recast as a function of $q\equiv\left|\bm{k}_{1}-\bm{k}_{2}\right|$
for $0.9k_{F}<|\bm{k}_{1}|,|\bm{k}_{2}|<1.1k_{F}$ are shown as blue
dots, and the values of $W_{\bm{k}_{1}\bm{k}_{2}}(\nu_{m})$ are shown
as red dots. The red solid lines show the fitting to the model Eq.~(\ref{eq:fitting}).
Right: residues of the fitting by using $M_{m}(q)$ shown in the inset
of Fig.~\ref{fig:W}.}
\end{figure}

The effective pairing interaction is shown in Fig.~\ref{fig:Scattering-amplitude-and}.

\subsubsection{Effective EPC matrix element }

From Fig.~\ref{fig:Scattering-amplitude-and}, we observe that the
effective pairing interaction vanishes at $q\rightarrow0$ and peaks
at $q\sim2k_{\mathrm{F}}$. Similar behaviors are also observed in
the density correlation function shown in Fig.~\ref{fig:Density-response-function}.
It suggests that the effective pairing interaction could be fitted
by the relation:
\begin{equation}
W_{\bm{k}_{1}\bm{k}_{2}}(\nu_{m})=\left|M_{m}(q)\right|^{2}\chi_{\mathrm{i}}(\bm{k}_{1}-\bm{k}_{2},\nu_{m})\label{eq:fitting}
\end{equation}
with $q\equiv\left|\bm{k}_{1}-\bm{k}_{2}\right|$, and $M_{m}(q)$
is interpreted as an effective EPC matrix element. We carry out the
fitting by assuming $M_{m}(q)=f_{m}(q)v_{\mathrm{ei}}(q)/\epsilon_{\mathrm{et}}(q)$
with $f_{m}(q)$ being a smooth function of $q$. The smooth function
is chosen to be an interpolation function of five control points at
$q/2k_{F}=\{0.2,0.5,0.75,0.8725,1\}$. The values of the scaling function
at these points are treated as fitting parameters. The resulting scaling
functions are shown in the inset of Fig.~\ref{fig:W}. The residues
of the fitting are shown in the right panel of Fig.~\ref{fig:Scattering-amplitude-and}.
It is evident that the relation fits the numerical results remarkably
well. 

A good fitting to Eq.~(\ref{eq:fitting}) is an indication of the
soundness of our numerical implementation and formalism. This is because
the existence of such a relation, while expected physically, is nowhere
near an obvious result from our formalism. Since it takes many intermediate
steps to obtain the effective pairing interaction numerically (see
Fig.~\ref{fig:Flowchart}), it is unlikely that an relation like
Eq.~(\ref{eq:fitting}) could emerge from numerical results had inconsistency/inaccuracy
existed in any of the intermediate steps.

\subsubsection{Interaction parameters\label{subsec:Interaction-parameters-and}}

\begin{figure}
\includegraphics[width=1\columnwidth]{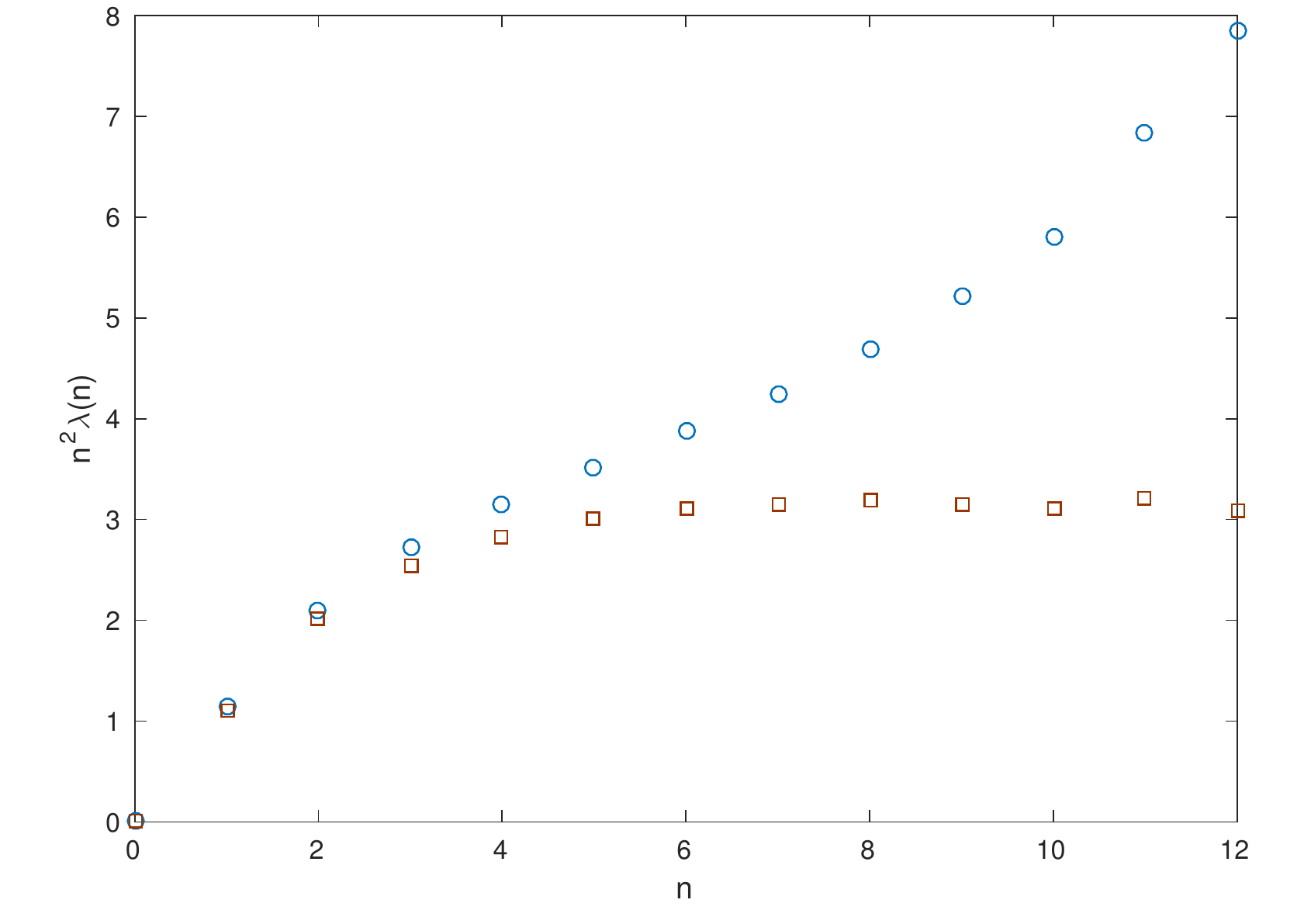}

\caption{\label{fig:lambdan}The values of $n^{2}\lambda(n)$ as a function
of $n$. Both the values determined from the original $W_{m}(q)$
(blue circles) and the oversampled $W_{m}^{(\mathrm{o})}(q)$ (red
squares) are shown.}
\end{figure}

We determine the interaction parameters $\lambda(n)$ by using Eq.~(\ref{eq:lambda})
with the effective pairing interaction determined by:
\begin{equation}
W_{m}^{(\mathrm{o})}(q)=\left|M_{m}(q)\right|^{2}\chi_{\mathrm{i}}^{(\mathrm{o})}(q,\nu_{m}),
\end{equation}
where $\chi_{\mathrm{i}}^{(\mathrm{o})}(q,\nu_{m})$ is the over-sampled
density correlation function determined in Sec.~\ref{subsec:Density-response-function}.
The over-sampling is necessary to eliminate the discretization errors
and to yield a correct asymptotic behavior of $\lambda(n)$ in the
large $n$ limit. Figure~\ref{fig:lambdan} shows the dependence
of $n^{2}\lambda(n)$ on $n$. Without the over-sampling, the values
of $n^{2}\lambda(n)$ keeps increasing with $n$. With the over-sampling,
$n^{2}\lambda(n)$ saturates at large $n$ as expected~\citep{allen1975}.
The saturation value yields an estimate of the average phonon frequency
$\bar{\omega}_{2}=\lim_{n\rightarrow\infty}(2\pi/\hbar\beta)\sqrt{n^{2}\lambda(n)/\lambda}$,
which enters into the Eliashberg equations by renormalizing $\mu^{\ast}$
when the equations are solved with a large-$n$ cutoff~\citep{allen1975}.
We take the recovery of the correct asymptotic behavior of $\lambda(n)$
as an indication of the soundness of our oversampling scheme discussed
in Sec.~\ref{subsec:Density-response-function}.

\subsection{Quasi-static approximation\label{subsec:Quasi-static-approximation}}

An important approximation we adopt to simplify the calculation is
the quasi-static approximation. Directly solving Eqs.~(\ref{eq:Gamma}--\ref{eq:W})
is numerically challenging. For instance, to solve Eq.~(\ref{eq:TRtau})
with a moderate setting of cutoffs, one may need $\sim10^{5}$ frequency-wave-vector
bases. Even worse, the solution may not have necessary accuracy because
it is difficult to evaluate $\bar{\mathcal{G}}(\omega_{n})$ accurately
in a PIMD simulation with a relatively small $N_{\mathrm{b}}$.

Fortunately, directly solving these time-dependent equations is not
necessary. We can exploit the fact that ions move much slowly than
electrons. As a result, the scattering potential $\hat{\mathcal{V}}(\tau)$
only has a few non-negligible low-frequency components. The resulting
$T$-matrix will be dominated by its frequency-diagonal components
$\hat{\mathcal{T}}_{nn}\equiv\hat{\mathcal{T}}(\omega_{n},\omega_{n})$,
and the amplitudes of off-diagonal components $\hat{\mathcal{T}}_{mn}$
with $|\omega_{m}-\omega_{n}|\gtrsim\omega_{\mathrm{ph}}$ are negligible.
For Eq.~(\ref{eq:TRtau}), we have:
\begin{align}
\hat{\mathcal{T}}_{mn} & =\hat{\mathcal{V}}_{m-n}+\frac{1}{\hbar}\hat{\mathcal{V}}_{m-n^{\prime}}\bar{\mathcal{G}}_{n^{\prime}}\hat{\mathcal{T}}_{n^{\prime}n}\\
 & \approx\hat{\mathcal{V}}_{m-n}+\frac{1}{\hbar}\hat{\mathcal{V}}_{m-n^{\prime}}\bar{\mathcal{G}}_{n}\hat{\mathcal{T}}_{n^{\prime}n},
\end{align}
where subscripts denote frequency components. The relative error induced
by the approximation is proportional to $|\omega_{n}-\omega_{n^{\prime}}|/\omega_{n}$,
and becomes negligible when $\omega_{n}\gg\left|\omega_{n}-\omega_{n^{\prime}}\right|\sim\omega_{\mathrm{ph}}$.

To solve the approximated equation, we choose $n$ to be a large integer
$N_{\mathrm{s}}$ such that $\omega_{\mathrm{ph}}\ll\omega_{N_{\mathrm{s}}}\ll\epsilon_{\mathrm{F}}/\hbar$.
The big disparity of the energy scales of electrons and phonons means
that one can always have such a choice. The equation can be conveniently
solved in the time-domain:
\begin{align}
\hat{\mathcal{T}}\left(\nu_{m}+\omega_{N_{\mathrm{s}}},\omega_{N_{\mathrm{s}}}\right) & =\frac{1}{\hbar\beta}\int_{0}^{\hbar\beta}\mathrm{d}\tau\hat{\mathcal{T}}_{N_{\mathrm{s}}}(\tau)e^{\mathrm{i}\nu_{m}\tau},\label{eq:TFourier}\\
\hat{\mathcal{T}}_{N_{\mathrm{s}}}(\tau) & =\hat{\mathcal{V}}(\tau)+\frac{1}{\hbar}\hat{\mathcal{V}}(\tau)\bar{\mathcal{G}}_{N_{\mathrm{s}}}\hat{\mathcal{T}}_{N_{\mathrm{s}}}(\tau),
\end{align}
where $\nu_{m}\equiv2\pi m/\hbar\beta$ is a Bosonic Matsubara frequency.
$\hat{\mathcal{T}}_{N_{\mathrm{s}}}(\tau)$ can be obtained for each
$\tau$ by solving an elastic Lippmann-Schwinger equation by treating
$\mathcal{V}(\tau)$ as if it is a static potential. We call the approximation
quasi-static approximation. By inserting the solution into Eq.~(\ref{eq:Gamma})
and averaging all ionic configurations, we can obtain a scattering
amplitude $\Gamma_{N_{\mathrm{s}}}(\nu_{m})\equiv\Gamma(\nu_{m}+\omega_{N_{\mathrm{s}}},\omega_{N_{\mathrm{s}}})$.

To solve the BS equation (\ref{eq:W}), we also apply the quasi-static
approximation. This is to approximate the equation as
\begin{multline}
\hat{W}(\nu_{m}+\omega_{N_{\mathrm{s}}},\omega_{N_{\mathrm{s}}})\approx\hat{\Gamma}_{N_{\mathrm{s}}}(\nu_{m})+\frac{1}{\hbar^{2}\beta}\sum_{m^{\prime}}\\
\hat{\Gamma}_{N_{\mathrm{s}}}(\nu_{m}-\nu_{m^{\prime}})\left|\hat{\bar{\mathcal{G}}}_{N_{\mathrm{s}}}\right|^{2}\hat{W}(\nu_{m^{\prime}}+\omega_{N_{\mathrm{s}}},\omega_{N_{\mathrm{s}}}).
\end{multline}
The resulting equation can then be solved in the time-domain in a
similar way like Eq.~(\ref{eq:TFourier}) (See Sec.~\ref{subsec:Bethe-Salpeter-equation}).

It is reasonable to expect that the effective interaction $W(\nu_{m}+\omega_{n},\omega_{n})$
is close to $W(\nu_{m}+\omega_{N_{\mathrm{s}}},\omega_{N_{\mathrm{s}}})$
as long as $\hbar\left|\omega_{n}-\omega_{N_{\mathrm{s}}}\right|\ll\epsilon_{\mathrm{F}}$:
\begin{equation}
W(\omega_{n}+\nu_{m},\omega_{n})\approx W(\nu_{m}+\omega_{N_{\mathrm{s}}},\omega_{N_{\mathrm{s}}}).
\end{equation}
It suggests that in the regime of interest with $|\omega_{n}|,|\omega_{n^{\prime}}|\ll\epsilon_{\mathrm{F}}/\hbar$,
the effective interaction $\hat{W}(\omega_{n},\omega_{n^{\prime}})$
is approximately a function of $\omega_{n}-\omega_{n^{\prime}}$,
and can be determined in the quasi-static limit.

We note that a similar approximation, i.e., treating $\hat{V}_{\mathrm{ei}}(\tau)$
as a static potential, is also adopted for PIMD simulations when determining
atomic forces. It is customary to call the approximation as an ``adiabatic
approximation''. Since the particular approximation does not prevent
us from determining the $\tau$-dependences of various physical quantities,
it does not affect the determination of EPC in an equilibrium system.
To avoid confusion, we call the approximation as a ``quasi-static
approximation'' since it is known that EPC is intrinsically non-adiabatic
and cannot be determined by an adiabatic approximation. The term ``adiabatic
approximation'' is reserved only for the Born-Oppenheimer approximation
employed by the classical molecular dynamics (see Sec.~\ref{subsec:Exact-decomposition-of}).

\subsection{Results\label{subsec:Results}}

\subsubsection{Metallic Hydrogen}

\begin{figure}
\includegraphics[width=1\columnwidth]{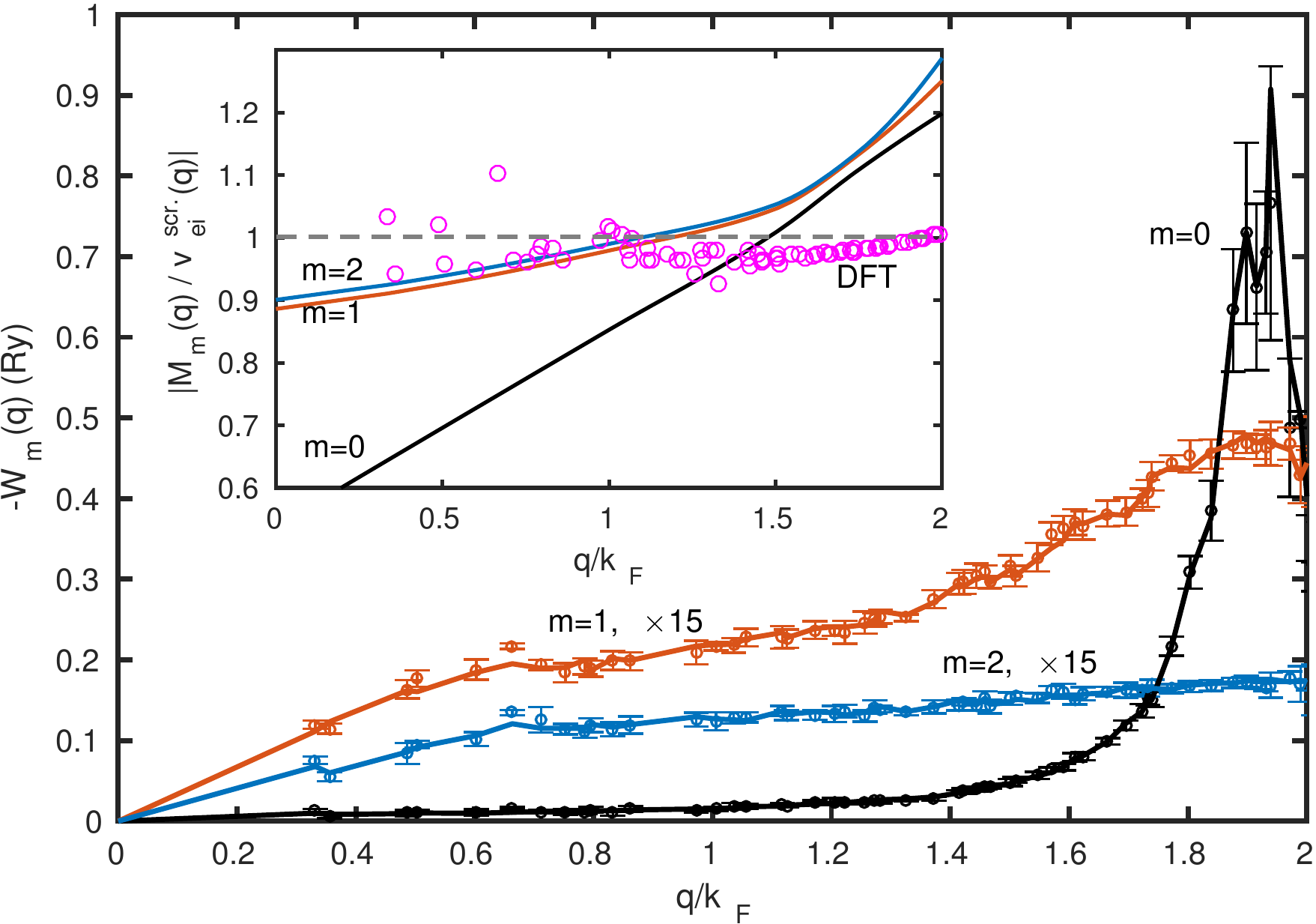}

\caption{\label{fig:W} (Color online) Effective pairing interaction $W_{m}(q)$
with $N_{\mathrm{s}}=16$, $m=0,1,2$, and $q\equiv|\bm{k}_{1}-\bm{k}_{2}|$
with $0.9k_{F}<|\bm{k}_{1}|,|\bm{k}_{2}|<1.1k_{F}$ for a metallic
hydrogen liquid at $P=700\,\mathrm{GPa}$ and $T=350\,\mathrm{K}$.
The calculation is based on a PIMD simulation of $200$ hydrogen atoms
and $6503$ samples of ion trajectories with the imaginary time discretized
to $24$ beads~\citep{chen2013}. The scatter points and error-bars
show the averages and standard deviations of $W_{m}(q)$ with same
$q$ but different $\bm{k}_{1}$'s and $\bm{k}_{2}$'s. The solid
lines show fittings to the model $W_{m}(q)\sim|M_{m}(q)|^{2}\chi_{\mathrm{i}}(\nu_{m},q)$.
Inset: the effective EPC matrix elements $M_{m}(q)$ (solid lines),
shown as ratios to the screened electron-ion potential $v_{\mathrm{ei}}^{\mathrm{scr.}}\equiv|v_{\mathrm{ei}}(q)/\epsilon_{\mathrm{et}}(q)$|.
The scatter points show the ratios between the effective ionic potential
determined from the DFT and that from the linear screening approximation,
averaged over $1455$ ionic configurations randomly sampled from the
PIMD simulation.}
\end{figure}

We summarize the result for the case of $P=0.7\,\mathrm{TPa}$ and
$T=350\,\mathrm{K}$ in Fig.~\ref{fig:W}. The effective pairing
interaction matrix elements $W_{\bm{k}_{1}\bm{k}_{2}}(\omega_{N_{\mathrm{s}}}+\nu_{m},\omega_{N_{\mathrm{s}}})$
are recast as a function $W_{m}(q)$ with $q\equiv|\bm{k}_{1}-\bm{k}_{2}|$
for $\bm{k}_{1}$'s and $\bm{k}_{2}$'s close to the Fermi surface.
The finiteness of the supercell of the PIMD simulation means that
$W_{m}(q)$ is only defined for a discrete set of $q$-values. To
this end, it is reasonable to assume that $W_{m}(q)$ is a smooth
function of $q$ and can be interpolated from the discrete set of
values. The effective EPC matrix element $M_{m}(q)$ is determined
and shown in the inset. As opposed to the earlier theoretical effort~\citep{jaffe1981},
the effective EPC matrix element can now be determined from first
principles.

We carry out PIMD simulations, determine the interaction parameters
and solve the Eliashberg equations for metallic hydrogen under a number
of pressures and at $T=350\,\mathrm{K}$ and $450\,\mathrm{K}$. The
results are summarized in Table~\ref{tab:lam}. Based on the results,
$T_{\mathrm{c}}$'s are estimated by linearly interpolating the values
of $\rho_{\mathrm{m}}$ between the two calculated temperatures. For
pressures ranging from $0.5\,\mathrm{TPa}$ to $1.5\,\mathrm{TPa}$,
they are close to $400\,\mathrm{K}$, well above the melting temperatures
determined in both Ref.~\citep{chen2013} and \citep{geng2015}.

\begin{table}
\begin{tabular*}{1\columnwidth}{@{\extracolsep{\fill}}crcccccc}
\toprule 
\multicolumn{2}{c}{$r_{\mathrm{s}}$} & $1.226$ & $1.197$ & $1.17$ & $1.149$ & $1.113$ & $1.049$\tabularnewline
\multicolumn{2}{c}{$P$ } & $0.5$ & $0.6$ & $0.7$ & $0.8$ & $1.0$ & $1.5$\tabularnewline
\midrule
\midrule 
\multirow{3}{*}{\begin{turn}{90}
$350\mathrm{K}$
\end{turn}} & $\lambda$ & $9.4(14)$ & $8.5(11)$ & $8.3(10)$ & $6.9(9)$ & $5.9(8)$ & $4.8(4)$\tabularnewline
 & $\bar{\omega}_{2}$ & $108(13)$ & $116(13)$ & $116(16)$ & $129(13)$ & $140(21)$ & $167(26)$\tabularnewline
 & $\rho_{\mathrm{m}}$ & $0.40(12)$ & $0.38(11)$ & $0.32(10)$ & $0.30(8)$ & $0.29(11)$ & $0.21(9)$\tabularnewline
\midrule
\multirow{3}{*}{\begin{turn}{90}
$450\mathrm{K}$
\end{turn}} & $\lambda$ & $7.4(13)$ & $7.2(10)$ & $7.2(11)$ & $6.1(9)$ & $5.2(7)$ & $4.3(3)$\tabularnewline
 & $\bar{\omega}_{2}$ & $121(27)$ & $120(18)$ & $121(19)$ & $147(19)$ & $156(20)$ & $179(20)$\tabularnewline
 & $\rho_{\mathrm{m}}$ & $\underline{0.06}(12)$ & $\underline{0.08}(9)$ & $\underline{0.31}(8)$ & $\underline{0.08}(10)$ & $\underline{0.12}(8)$ & $\underline{0.15}(6)$\tabularnewline
\midrule
\midrule 
\multicolumn{2}{c}{$T_{\mathrm{c}}\,(\text{K})$} & $437(27)$ & $433(22)$ & $401(15)$ & $429(25)$ & $421(24)$ & $408(19)$\tabularnewline
\bottomrule
\end{tabular*}

\caption{\label{tab:lam} Mass enhancement factor $\lambda\equiv\lambda(0)$,
average phonon frequency $\bar{\omega}_{2}$ (in meV), and the maximal
eigenvalue $\rho_{\mathrm{m}}$ of the linearized Eliashberg equations,
calculated for a number of temperatures and pressures $P$ (in $\mathrm{TPa}$).
$\bar{\omega}_{2}$ is estimated by applying the asymptotic relation
$\bar{\omega}_{2}=\lim_{n\rightarrow\infty}(2\pi/\hbar\beta)\sqrt{n^{2}\lambda(n)/\lambda}$~\citep{allen1975}.
Negative values are indicated by underlined numbers. Numerical uncertainties
are estimated by shifting the values of $W_{m}(q)$ up/down by a standard
deviation simultaneously for all the discrete $q$-values, and indicated
in parentheses. $T_{\mathrm{c}}$ is estimated from the linear interpolation
of $\rho_{\mathrm{m}}$. We adopt $\mu^{\ast}\approx0.089$ for the
Coulomb pseudo-potential~\citep{mcmahon2011}. The band renormalization
is found to be negligible for metallic hydrogen. }
\end{table}

\subsubsection{Metallic deuterium and isotope effect}

\begin{figure}
\includegraphics[width=0.95\columnwidth]{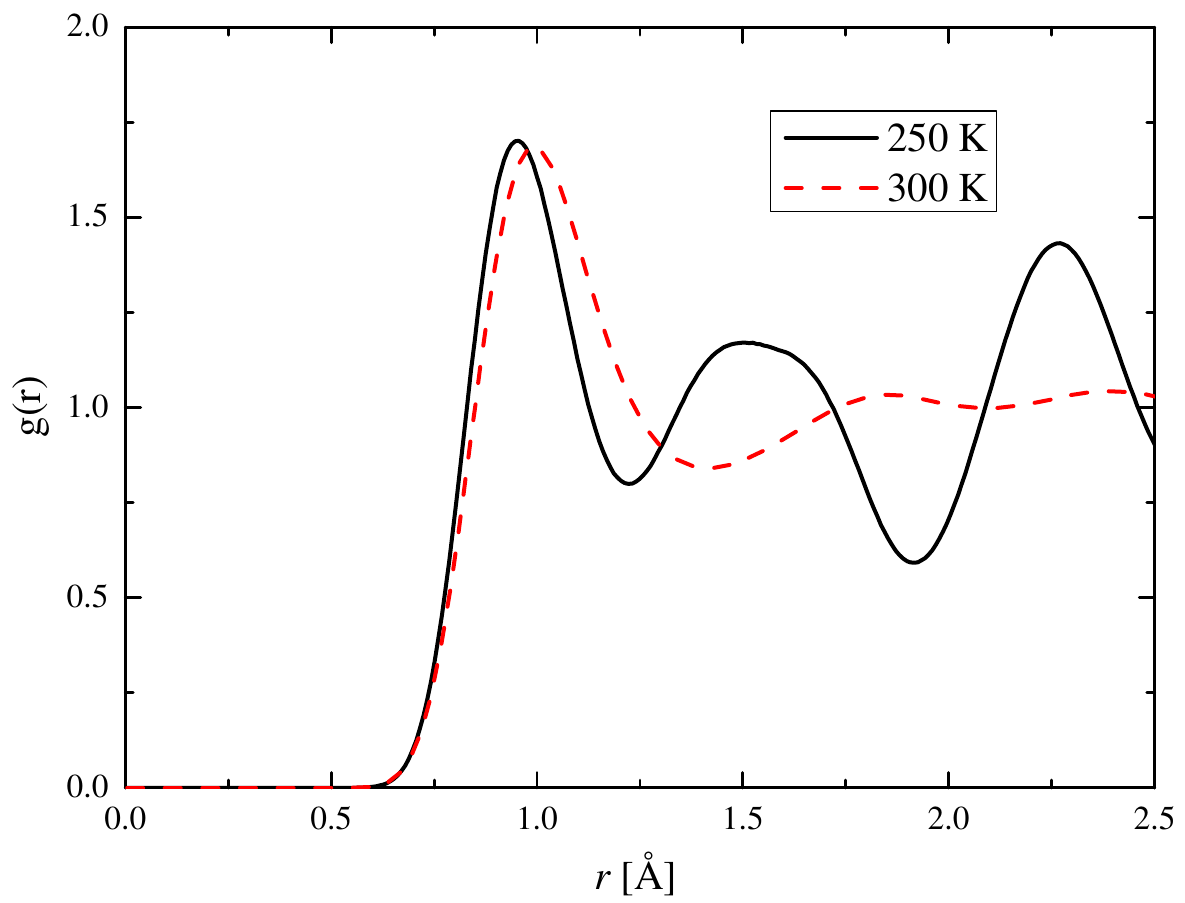}

\caption{\label{fig:Radial-pair-distribution}Radial pair distribution function
$g(r)$ for deuterium at $P=1\,\mathrm{TPa}$. The functions for both
$T=250\,\mathrm{K}$ (black solid line) and $T=300\,\mathrm{K}$ (red
dashed line) are shown.}
\end{figure}

A test to our approach is to see whether or not it predicts the isotope
effect as expected. For the purpose, we carry out PIMD simulations
for metallic deuterium at $P=1\,\mathrm{TPa}$. The simulations are
performed at 250, 300 and 350 K for a time interval of $5\,\mathrm{ps}$.
The radial pair distribution function (RDF) $g(r)$ is calculated.
As shown in Fig.~\ref{fig:Radial-pair-distribution}, the RDF for
$T=250\,\mathrm{K}$ shows sharp peaks, which indicates a solid state.
At $T=300\,\mathrm{K}$, the sharp peaks after the first one become
broad humps, which suggests a liquid state. We thus conclude that
the melting temperature for deuterium at $P=1\,\mathrm{TPa}$ is between
$250\,\mathrm{K}$ and $\mathrm{300\,\mathrm{K}}$.

\begin{figure}
\includegraphics[width=1\columnwidth]{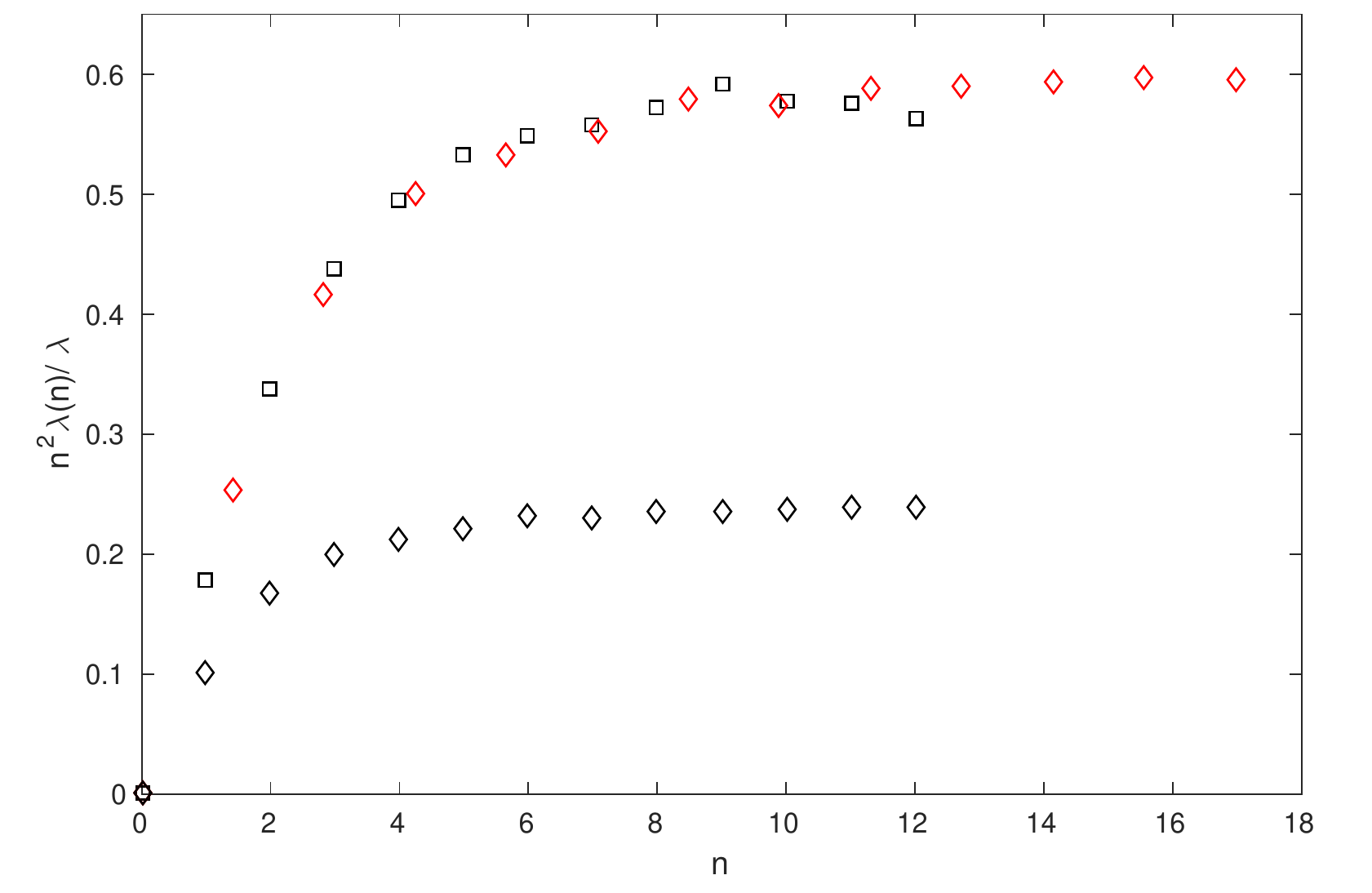}

\caption{\label{fig:DHlambdan}$n^{2}\lambda(n)/\lambda$ vs. $n$ for both
metallic hydrogen (black squares) and metallic deuterium (black diamonds)
at $T=350\,\mathrm{K}$ and $P=1\,\mathrm{TPa}$. For comparison,
the deuterium data are also shown scaled (red diamonds) by factors
$\sqrt{2}$ and 2.5 along the $x$- and $y$-directions, respectively.}
\end{figure}

We carry out analyses for the PIMD data. Figure \ref{fig:DHlambdan}
shows a comparison between results for hydrogen and deuterium. For
the relation of $n^{2}\lambda(n)/\lambda$ vs. $n$ shown, the isotope
effect predicts that the two traces would collapse into one if the
deuterium data are scaled by factors $\sqrt{2}$ and 2 along the $x$-
and $y$-directions, respectively. In the plot, we see that the respective
factors are $\sqrt{2}$ and $2.5$. For deuterium, we determine $\hbar\bar{\omega}_{2}\approx93\pm9\,\mathrm{meV}$,
while for hydrogen $\hbar\bar{\omega}_{2}\approx140\pm21\,\mathrm{meV}$
(Table \ref{tab:lam}). The ratio between the two is also close to
$\sqrt{2}$ predicted by the isotope effect.

To estimate $T_{\mathrm{c}}$, we analyze PIMD data at $300\,\mathrm{K}$
and $350\,\mathrm{K}$. The maximal eigenvalues of the Eliashberg
equations are $-0.06\pm0.08$ and $-0.30\pm0.05$, respectively. It
indicates that $T_{\mathrm{c}}$ is lower than $300\,\mathrm{K}$.
An estimate by \emph{extrapolation} yields $T_{\mathrm{c}}\approx288\,\mathrm{K}$
for deuterium, close to the prediction of the isotope effect $421\,\mathrm{K}/\sqrt{2}\approx298\,\mathrm{K}$.

\section{Summary\label{sec:Summary}}

In summary, we have developed a non-perturbative approach for calculating
$T_{\mathrm{c}}$'s of liquids. The approach could be implemented
as a first-principles tool of searching for EPC superconductivity
in liquids. It predicts that a metallic hydrogen liquid is a superconducting
liquid at room temperature. Experimentally, it implies that metallic
hydrogen could be detected by measuring the diamagnetism induced by
the Meissner effect.

Our approach can also be applied to more general systems such as (anharmonic)
solids. The numerical implementation shown in this paper, however,
is only applicable for metallic hydrogens for which the linear screening
approximation is satisfactory. For the more general systems, it is
desirable to eliminate the linear screening approximation and determine
the ionic fields from first principles. This is still a work ongoing.
\begin{acknowledgments}
We thank Xiaowei Zhang for pointing out that an equation like Eq.~(\ref{eq:imsigma})
also arises in disordered electron systems~\citep{vanoosten1985}.
This work is supported by National Basic Research Program of China
(973 Program) Grant No. 2015CB921101, 2016YFA0300900, and National
Natural Science Foundation of China Grant No. 11325416, 11774003.
\end{acknowledgments}

\bibliographystyle{apsrev4-1}
\bibliography{SCLiquids}

\end{document}